\documentclass[lettersize,journal]{IEEEtran}
\usepackage[dvipsnames,svgnames,x11names]{xcolor}
\usepackage{algorithmic}
\usepackage{algorithm}
\usepackage{array}
\usepackage[caption=false,font=normalsize,labelfont=sf,textfont=sf]{subfig}
\usepackage{textcomp}
\usepackage{stfloats}
\usepackage{bibspacing}
\usepackage{url}
\usepackage{verbatim}
\usepackage{cite}
\usepackage{multirow}
\usepackage{bbding}
\usepackage{booktabs}
\usepackage[utf8]{inputenc}
\usepackage{makecell}
\usepackage{tipx}

\newcommand{\xmark}{\ding{55}}

\usepackage{color}
\usepackage{hhline}
\usepackage{pifont}
\usepackage{bm}
\usepackage{amsmath,amsfonts,amssymb,mathtools}

\usepackage{graphicx,float}

\usepackage{etoolbox}  
\makeatletter  
\patchcmd{\@footnotetext}{\footnotesize}{\scriptsize}{}{}  
\makeatother 

\usepackage[normalem]{ulem}

\begin{document}
\bstctlcite{IEEEexample:BSTcontrol}
\title{
Self-supervised ASR Models and Features For Dysarthric and Elderly Speech Recognition}

\author{Shujie Hu, Xurong Xie, Mengzhe Geng, Zengrui Jin, Jiajun Deng, Guinan Li, Yi Wang, \\ Mingyu Cui, Tianzi Wang, Helen Meng~\IEEEmembership{Fellow,~IEEE}, Xunying Liu~\IEEEmembership{Member,~IEEE}
    \thanks{Shujie Hu, Mengzhe Geng, Zengrui Jin, Jiajun Deng, Guinan Li, Yi Wang, Mingyu Cui, Tianzi Wang are with the Chinese University of Hong Kong, China (email: \{sjhu,mzgeng,zrjin,jjdeng,gnli,mycui,twang\}@se.cuhk.edu.hk).\\
    \indent Xurong Xie is with Institute of Software, Chinese Academy of Sciences, Beijing, China (email: xurong@iscas.ac.cn). \\
    \indent Helen Meng is with the Chinese University of Hong Kong, China (email: hmmeng@se.cuhk.edu.hk).\\
    \indent Xunying Liu is with the Chinese University of Hong Kong, China (email: xyliu@se.cuhk.edu.hk). \\
    \indent Xunying Liu and Xurong Xie are the corresponding authors.
    }
}

\maketitle

\begin{abstract}
Self-supervised learning (SSL) based speech foundation models have been applied to a wide range of ASR tasks. However, their application to dysarthric and elderly speech via data-intensive parameter fine-tuning is confronted by in-domain data scarcity and mismatch. To this end, this paper explores a series of approaches to integrate domain fine-tuned SSL pre-trained models and their features into TDNN and Conformer ASR systems for dysarthric and elderly speech recognition. These include: a) input feature fusion between standard acoustic frontends and domain fine-tuned SSL speech representations; b) frame-level joint decoding between TDNN systems separately trained using standard acoustic features alone and those with additional domain fine-tuned SSL features; and c) multi-pass decoding involving the TDNN/Conformer system outputs to be rescored using domain fine-tuned pre-trained ASR models. In addition, fine-tuned SSL speech features are used in acoustic-to-articulatory (A2A) inversion to construct multi-modal ASR systems. Experiments are conducted on four tasks: the English UASpeech and TORGO dysarthric speech corpora; and the English DementiaBank Pitt and Cantonese JCCOCC MoCA elderly speech datasets. 
The TDNN systems constructed by integrating domain-adapted HuBERT, wav2vec2-conformer or multi-lingual XLSR models and their features consistently outperform the standalone fine-tuned SSL pre-trained models. These systems produced statistically significant WER or CER reductions of \textbf{6.53\%}, \textbf{1.90\%}, \textbf{2.04\%} and \textbf{7.97\%} absolute (\textbf{24.10\%}, \textbf{23.84\%}, \textbf{10.14\%} and \textbf{31.39\%} relative) on the four tasks respectively. 
Consistent improvements in Alzheimer's Disease detection accuracy are also obtained using the DementiaBank Pitt elderly speech recognition outputs.
\end{abstract}


\vspace{-0.3cm}
\section{Introduction}
\IEEEPARstart{D}ESPITE the rapid progress of automatic speech recognition (ASR) technologies targeting normal speech in recent decades 
\cite{peddinti2015time, gulati2020conformer}
, accurate recognition of dysarthric and elderly speech remains highly challenging tasks to date \cite{christensen2012comparative, christensen2013combining, sehgal2015model, yu2018development, hu2019cuhk, liu2020exploiting, ye2021development, jin2022personalized, geng2022speaker, geng23b_interspeech, joy2018improving, hu2022exploring}. Dysarthria is a common type of speech disorder caused by a wide spectrum of motor control conditions including cerebral palsy, amyotrophic lateral sclerosis, stroke and brain injuries. 
In addition, neurocognitive disorders, such as Alzheimer’s disease (AD),
are often found among the elderly with speech and language impairments \cite{fraser2016linguistic, alzheimer20192019}.  
\par
ASR technologies tailored to dysarthric and elderly users' needs not only improve their quality of life but also enable large-scale automatic early diagnosis of neurocognitive impairment, such as AD \cite{rohanian21_interspeech, ammar2018speech, li2021comparative}. To this end, recently there has been increasing interest in developing ASR technologies for dysarthric \cite{christensen2013combining, vachhani2017deep, kim2018dysarthric, joy2018improving, hu2022exploiting, lin20n_interspeech, xiong2020source, liu2021recent, xie21b_interspeech, takashima2020two, hermann2020dysarthric, wang21_interspeech, macdonald21_interspeech, wang2021improved, green21_interspeech, 9747516, shor19_interspeech, yue22_interspeech, prananta22_interspeech, violeta2022investigating, bhat22_interspeech, hernandez22_interspeech, tomanek2023analysis, yue2022acoustic, jin2023adversarial, 9746585} and elderly users \cite{ye2021development, vipperla2010ageing, rudzicz2014speech, zhou16_interspeech, hu2022exploitinguti, wang22k_interspeech, konig2018fully, toth2018speech, pan21c_interspeech, 9747167, 9747856, 10095593, yang22k_interspeech, ablimit22_interspeech, 10096770}.
\vspace{-0.7cm}
\subsection{Dysarthric and Elderly Speech Recognition Prior to SSL Pre-trained Speech Models}
\label{Sec_IA}
Dysarthric and elderly speech bring considerable challenges to current ASR technologies primarily targeting normal speech recorded from healthy, non-aged users. These include: \textbf{a) large mismatch against normal speech} due to motor control conditions
and aging, e.g., articulation imprecision, decreased speech volume and clarity, and increased disfluency; \textbf{b) data scarcity} due to the difficulty in data collection from these speakers with physical disabilities and mobility issues; and \textbf{c) large speaker-level diversity}. A set of publicly available dysarthric and elderly speech corpora are shown in Table \ref{table_dataset}.

\begin{table}[h]
\vspace{-0.3cm}
\centering
\caption{Description of publicly available dysarthric and elderly speech corpora for English (ENG.) and Cantonese (CAN.). ``Lang.'' and ``Vocab.'' stand for language and the number of words in vocabulary respectively.}
\scalebox{0.8}{
\begin{tabular}{c|c|c|c|c|c} 
\hline\hline
Corpus            & Type                        & Lang.                 & \# Hour & \# Speaker & \# Vocab.  \\ 
\hline\hline
UASpeech \cite{kim2008dysarthric}          & \multirow{2}{*}{Dysarthric} & \multirow{2}{*}{Eng.} & 102.7    & 29       & 455 words       \\
TORGO \cite{rudzicz2012torgo}            &                             &                       & 15       & 15       & 1573 words      \\ 
\hline
DementiaBank Pitt \cite{becker1994natural} & \multirow{2}{*}{Elderly}    & Eng.                  & 33.1     & 688      & 3.8k words      \\
JCCOCC MoCA \cite{xuspeaker}      &                             & Can.                  & 52       & 369      & 610k words      \\
\hline\hline
\end{tabular}
}
\label{table_dataset}
\end{table}
In order to address the data scarcity issue, one major area of prior researches focused on data augmentation techniques. Speed and temporal perturbation were employed in \cite{geng2022speaker, liu2021recent, ye2021development, hu2022exploitinguti, verhelst1993overlap, kanda2013elastic, ko2015audio}. In addition to speaker-independent perturbation of limited in-domain dysarthric or elderly speech only using fixed perturbation factors such as \{0.9, 1, 1.1\}, more powerful speaker-dependent data augmentation approaches were also developed \cite{geng2022speaker, liu2021recent, ye2021development, hu2022exploitinguti, jin2021adversarial, jin2023adversarial, xiong2019phonetic}. 
Adversarial learning \cite{jin2021adversarial, bhat22_interspeech, jin2023adversarial, harvill2021synthesis, jiao2018simulating, prananta2022effectiveness}, voice conversion \cite{huang2022towards, huang2021preliminary} and text-to-speech synthesis \cite{wang2023duta} based data augmentation approaches designed for dysarthric or elderly speech were also developed. 
Another major area of prior researches addressing both the issues of data scarcity and mismatch against normal speech focused on domain adapting general purpose ASR systems that are trained using large quantities of out-of-domain healthy and non-aged speech data \cite{deng21d_interspeech, ye2021development, yu2018development, takashima2020two, wang22k_interspeech, wang2021improved, shor19_interspeech, green21_interspeech, 9747516}.
To model the diversity of dysarthric or elderly speakers, a range of speaker adaptation techniques were studied. These include, but are not limited to, direct parameter fine-tuning \cite{9053725, shor19_interspeech}; using compact learning hidden unit contributions (LHUC) \cite{liu2021recent, geng23_interspeech}; and spectro-temporal deep features \cite{geng2022speaker}. In addition, the incorporation of visual \cite{liu2021recent, liu2020exploiting, liu2019exploiting, salama2014audio, takashima2017audio} and articulatory movement features \cite{hu2022exploitinguti, hu2022exploiting, maharana2021acoustic} allows multi-modal dysarthric and elderly speech recognition systems to be constructed. In these systems \cite{liu2021recent, hu2022exploitinguti, hu2022exploiting}, frame-level joint decoding is used to fuse audio-only and multi-modal ASR systems, before cross-system multi-pass decoding \cite{hu2022exploitinguti} being further used.

\vspace{-0.4cm}
\subsection{Dysarthric and Elderly Speech Recognition using SSL Pre-trained Speech Models}
\label{Sec_IB}
The recent emergence of self-supervised learning (SSL) speech foundation models \cite{baevski2020wav2vec, chen2022wavlm, hsu2021hubert, pmlr-v162-baevski22a, babu2021xls} provides a new paradigm to address the above data scarcity and domain mismatch problems.
These speech foundation models are SSL pre-trained on large quantities of unlabelled data. They have been successfully applied to normal speech processing tasks including ASR \cite{baevski2020wav2vec, chen2022wavlm, hsu2021hubert, pmlr-v162-baevski22a, conneau2020unsupervised}, speech emotion recognition \cite{pepino2021emotion}, speaker recognition \cite{9746952}, voice conversion \cite{9746430} and speech synthesis \cite{ni22_interspeech}. In addition, SSL speech representations are robust to domain mismatch \cite{hernandez22_interspeech,9746250}. 
\par
In contrast, more limited prior researches on applying SSL pre-trained models to dysarthric and elderly speech have been conducted. Wav2vec2.0 \cite{baevski2020wav2vec} and WavLM \cite{chen2022wavlm} models were applied to Japanese electrolaryngeal and English dysarthric speech in \cite{violeta2022investigating}, where an overall word error rate (WER) of 51.8\% was reported on the benchmark UASpeech \cite{kim2008dysarthric} task. Speaker adaptation of the Wav2vec2.0 model using fMLLR and x-vectors during fine-tuning for dysarthric speech recognition was investigated in \cite{baskar2022speaker}. Cross-lingual SSL-based dysarthric speech recognition was studied in \cite{hernandez22_interspeech}, where SSL speech representations were extracted from fine-tuned Wav2vec2.0 \cite{baevski2020wav2vec}, cross-lingual XLSR \cite{conneau2020unsupervised} and HuBERT \cite{hsu2021hubert} models before being fed into the Conformer based ASR systems. Incorporating Wav2vec2.0 models and features into hybrid TDNN and end-to-end Conformer systems was proposed in \cite{hu2022exploring}, including input feature fusion, frame-level joint decoding and cross-system multi-pass rescoring. Speech impairment severity was incorporated by adding cross-entropy based severity prediction error in Wav2vec2.0 fine-tuning \cite{geng23b_interspeech}. Wav2vec2.0 embedding features were used to learn dysarthric speech characteristics in the VAE-GAN based personalized disordered speech augmentation approaches \cite{jin2023adversarial}. The authors in \cite{yu2023multi} proposed to use the AV-HuBERT model to fuse audio and visual modalities to improve the performance of dysarthric speech recognition. A WER of 63.98\% on the very low intelligibility group of the benchmark UASpeech dataset was reported. In addition to English, the Wav2vec2.0 model and cross-lingual XLSR model were also evaluated on Dutch dysarthric home-automation data in \cite{wang2023benefits}. Speaker-dependent fine-tuning of SSL pre-trained ASR models for Dutch dysarthric speech was studied in \cite{matsushima2022dutch}. 

\vspace{-0.4cm}
\subsection{Key Research Problems and Methodology Design}
\textbf{1) Pre-trained ASR performance disparity and fairness:} Data-intensive fine-tuning a large number of pre-trained ASR model parameters on limited impaired or elderly speech data rapidly leads to poor generalization. This issue is further exacerbated when limited training data provides insufficient coverage. 
For example, approximately 39\% of words in the benchmark UASpeech test set do not occur in the training data. Under such conditions, current mainstream end-to-end ASR systems including SSL pre-trained models have been found to produce large performance disparity on two fronts: \textbf{a)} between seen and unseen words in the often very limited dysarthric speech \cite{liu2021recent, geng23b_interspeech}; and \textbf{b)} between impaired speakers with high and very low speech intelligibility \cite{wang2021improved, geng23b_interspeech}.

\textbf{2) Use of SSL speech foundation models:} Their application to dysarthric and elderly speech needs to account for the underlying issues over data scarcity and domain mismatch. To this end, alternative approaches that can effectively exploit SSL pre-trained ASR models and feature representations, while exhibiting less performance fragility over insufficient data coverage \cite{liu2021recent, geng23b_interspeech} than using the fine-tuned ASR models alone, need to be investigated.

\textbf{3) Articulatory features generation and SSL models:} Articulatory features are inherently robust to be acoustic signal perturbation. They have been successfully applied to normal and pathological speech \cite{rudzicz2010articulatory, gonzalez2017direct, xiong2018deep, yilmaz2019articulatory} recognition. However, such scarce and specialist data has been traditionally collected mainly from healthy, non-aged English speakers \cite{cai2011recognition, eshky2019ultrasuite, ribeiro2021tal, rudzicz2012torgo, richmond11_interspeech, wrench2000multi}. This hinders their application to dysarthric and elderly speech across multiple languages. To this end, acoustic-to-articulatory (A2A) inversion techniques further empowered by domain and language invariant SSL pre-trained speech representations need to be developed.

\textbf{4) Multi-faceted user applications and evaluation metrics:} Advancement of dysarthric and elderly speech recognition technologies in recent decades have broadened their scope of application beyond ASR-based assistive technology to aid communication and improve quality of life for such users \cite{christensen2012comparative, christensen2013combining, sehgal2015model, yu2018development, hu2019cuhk, liu2020exploiting, jin2022personalized, geng2022speaker, geng23b_interspeech, joy2018improving, hu2022exploring, jin2023adversarial, wang2023hyper, liu2021recent, hu2022exploitinguti,jin2021adversarial}. Aging presents enormous challenges to health care worldwide. Neurocognitive disorders, such as Alzheimer’s disease (AD), are often found among older adults and manifest themselves in speech and language impairments \cite{fraser2016linguistic, alzheimer20192019}. Dysarthric and elderly speech recognition forms the core technology to facilitate fully automated, large-scale, less intrusive and low-cost neurocognitive disorders screening among the aging population \cite{ye2021development, wang22l_interspeech, wang22k_interspeech, hu2022exploring,rohanian21_interspeech, ammar2018speech, li2021comparative, wang2023exploiting, rudzicz2014speech, zhou16_interspeech, ablimit22_interspeech, 9747856, pan21c_interspeech}. To this end, dysarthric and elderly speech recognition technologies tailored for such a wider range of medical domain applications need to be evaluated not only using ASR word error rate, but also additional metrics that are relevant to the specific task such as AD detection.
\par
In order to address the above issues, this paper explores a series of techniques to integrate state-of-the-art mono-lingual and multi-lingual SSL pre-trained speech foundation models and their features into hybrid TDNN \cite{peddinti2015time} and Conformer \cite{gulati2020conformer} ASR systems for dysarthric and elderly speech recognition. \textbf{1)} We aim to exploit the diversity and complementarity among them, and to improve the generalization performance on unseen and poorly covered words as well as on the most challenging dysarthric speech data of very low intelligibility.
\textbf{2)} These include:
\textbf{a) input feature fusion} between standard acoustic frontends and fine-tuned SSL speech representations; \textbf{b) time-synchronous frame-level joint decoding} \cite{liu2021recent, swietojanski2013revisiting, hu2022exploiting} of TDNN systems separately trained using standard acoustic features alone and those with additional fine-tuned SSL representations in {\bf a)}; and \textbf{c) cross-system multi-pass decoding} \cite{cui2022two} involving the TDNN or Conformer systems' N-best outputs to be rescored using domain-adapted pre-trained models.
\textbf{3)} Finally, domain-adapted SSL representations are utilized in acoustic-to-articulatory (A2A)
inversion  \cite{hu2022exploitinguti} to produce ultrasound tongue imaging (UTI) \cite{ribeiro2021tal} articulatory movement features based multi-modal ASR systems. \textbf{4)} The performance of the ASR system will be evaluated not only using ASR word error rate, but also additional metrics that are relevant to the specific task such as AD detection.

\par
Experiments are conducted on four tasks: the English UASpeech \cite{kim2008dysarthric} and TORGO \cite{rudzicz2012torgo} dysarthric speech corpora; the English DementiaBank Pitt \cite{becker1994natural} and Cantonese JCCOCC MoCA \cite{xuspeaker} elderly speech datasets. Among these, the UASpeech and DementiaBank Pitt corpora are respectively the largest publicly available datasets for dysarthric speech and elderly speech. The TDNN systems constructed by integrating domain-adapted HuBERT, wav2vec2-conformer or multi-lingual XLSR models and their features consistently outperform the standalone domain fine-tuned SSL pre-trained models by statistically significant WER or character error rate (CER) reductions of \textbf{6.53\%}, \textbf{1.90\%}, \textbf{2.04\%} and \textbf{7.97\%} absolute (\textbf{24.10\%}, \textbf{23.84\%}, \textbf{10.14\%} and \textbf{31.39\%} relative) on the UASpeech, TORGO, DementiaBank Pitt and JCCOCC MoCA corpora respectively. 

\vspace{-0.3cm}
\subsection{Main Contributions}
\textbf{1)} To the best of our knowledge, this paper presents the first work to systematically investigate the generalization capability of mono-lingual and multi-lingual SSL pre-trained ASR models for dysarthric and elderly speech recognition tasks. Their generalization is measured against two objectives: \textbf{a)} minimizing the overall WER with a focus on the most challenging portions of elderly or dysarthric speech data with very low (VL) intelligibility, as current ASR performance on such data with mild (M) and high (H) intelligibility are comparable to that for normal, healthy and non-aged speakers; and \textbf{b)} improving model generalization and ``fairness'' in performance on unseen or poorly covered words that are often not sufficiently modeled for under-resourced pathological and medical speech data. In contrast,  previous works \cite{hernandez22_interspeech, violeta2022investigating, baskar2022speaker, yu2023multi} only analyzed the performance disparity across different speech impairment severity subsets. The generalization to unseen or rare words and sentences that are insufficiently covered in training was not studied.
\par
\textbf{2)} To address such generalization issues, this paper proposes novel approaches to integrate mono-lingual and multi-lingual SSL pre-trained speech models and their features into back-end ASR systems constructed using in-domain dysarthric or elderly speech data. Drawing strengths from both, the lowest published WERs of \textbf{20.56\%} (\textbf{50.70\%} on very low intelligibility, \textbf{34.28\%} on unseen words)\footnote{All the UASpeech experiments of this paper follow the University of Sheffield defined block based training and evaluation data partition \cite{christensen2012comparative, christensen2013combining, sehgal2015model, xiong2019phonetic}: all the data of Block 1 and 3 are used for training while the Block 2 dysarthric data serves as the test set.} and \textbf{18.07\%} are obtained on the benchmark UASpeech test set of 16 dysarthric speakers and the DementiaBank Pitt evaluation set of 48 elderly subjects. In contrast, prior researches focused on using stand-alone domain-adapted pre-trained models \cite{baskar2022speaker, violeta2022investigating, yu2023multi}.
\par
\textbf{3)} This work presents the novel use of SSL-based speech representations for cross-domain and cross-lingual A2A inversion. In contrast, prior researches on dysarthric and elderly speech are conducted predominantly using non-SSL based domain adaptation methods in  A2A inversion \cite{hu2022exploiting, hu2022exploitinguti}.
\par
\textbf{4)} The efficacy of our SSL pre-trained models and features integration approaches also leads to higher AD detection accuracy.
Using ASR outputs of the final systems to extract textual features from AD assessment speech recordings, a state-of-the-art speech recognition based AD detection mean accuracy of \textbf{83.94\%} (with the standard deviation of 4.58\%, and the best score of \textbf{95.83\%}) is obtained on DementiaBank Pitt evaluation set (ADReSS2020 \cite{luz2020alzheimer}) of 48 elderly speakers. 
\par
The rest of this paper is organized as follows. SSL pre-trained foundation speech models are reviewed in Sec. \ref{sec:ssl_models}. Sec. \ref{sec:a2a_inversion} introduces SSL speech representations based cross-domain and cross-lingual A2A inversion. A series of novel approaches to integrate pre-trained ASR models and their features into in-domain dysarthric and elderly speech data constructed TDNN and Conformer ASR systems are proposed in Sec. \ref{sec:sys_integrate}. A set of implementation issues affecting the performance of the baseline fine-tuned SSL pre-trained models and their integration with in-domain data trained ASR systems are discussed in Sec. \ref{sec:implementation}. Sec. \ref{sec:experiments} presents the experimental results and analysis. Sec.\ref{sec:discusion_conclusion} draws the conclusion and discusses future works.

\vspace{-0.1cm}
\section{SSL Pre-trained ASR Models}\label{sec:ssl_models}

This section gives an overview of SSL pre-trained ASR models, including Wav2vec2.0 \cite{baevski2020wav2vec}, HuBERT \cite{hsu2021hubert}, WavLM \cite{chen2022wavlm} and Data2vec \cite{pmlr-v162-baevski22a} models.
\vspace{-0.3cm}
\subsection{Pre-trained Wav2vec2.0 Model}
Wav2vec2.0 is a pre-trained model that jointly learns latent contextualized speech representations and an inventory of discretized latent speech units serving as the pseudo-labels. Contrastive learning based SSL pre-training is performed by distinguishing the target from distractor pseudo-labels.
\par
\noindent
\textbf{Model Architecture}: Wav2vec2.0 consists of three components, including 1) a multi-layer CNN-based feature encoder which encodes raw speech audio input $\boldsymbol{\mathcal{X}}$ into continuous speech representations $\boldsymbol{z}_t \in \boldsymbol{Z}$ with a stride of 20 ms and a receptive field of 25 ms; 2) a $L$-layers transformer-based context network producing contextual representations $\boldsymbol{c}_t \in \boldsymbol{C}$ over a sequence of randomly masked feature encoder outputs; and 3) a quantization module generating discrete speech units $\boldsymbol{q}_t \in \boldsymbol{Q}$ as \textbf{pseudo-labels} for SSL based pre-training.
\par
\noindent
\textbf{SSL Pseudo-labels}: After mapping the feature encoder output $\boldsymbol{z}_t$ to the logits $[\boldsymbol{l_1}, \boldsymbol{l_2},...,\boldsymbol{l_G}] \in {\mathbb{R}^{G\times V}}$,
the best code is chosen among all the $V$ entries of each codebook by Gumbel-softmax re-parameterization.
The discrete quantized unit $\boldsymbol{q}_t$ is obtained by applying a linear transformation to the concatenated $G$ codes, before serving as the pre-training pseudo-labels.
\par
\noindent
\textbf{SSL Criterion}: Wav2vec2.0 is pre-trained via an interpolation between a contrastive task $\mathcal{L}_c$ and a diversity task $\mathcal{L}_d$. 

\vspace{-0.5cm}
\begin{equation}
\resizebox{0.9\linewidth}{!}{
\begin{math}
     \setlength{\abovedisplayskip}{2pt}
    \setlength{\belowdisplayskip}{2pt}\mathcal{L}_{w2v, t}=\underbrace{-\log \frac{\exp \left(\operatorname{sim}\left(\boldsymbol{c}_t, \boldsymbol{q}_t\right) / \kappa\right)}{\sum\limits_{\boldsymbol{\tilde{q}} \in \boldsymbol{Q}_t} \exp\left(\operatorname{sim}\left(\boldsymbol{c}_t, \boldsymbol{\tilde{q}}\right) / \kappa\right)}}_{\text{\normalsize$\mathcal{L}_{c}$:} {\text{\normalsize ~Contrastive task}}} + \underbrace{\frac{\alpha}{G V} \sum\limits^G_{g=1} \sum\limits_{v=1}^V \bar{l}_{g, v} \log \bar{l}_{g, v}}_{\text{\normalsize$\mathcal{L}_d$:}{\text{\normalsize ~Diversity task}}}
\end{math}
}
\label{eqn:wav2vec}
\end{equation}
where $\operatorname{sim}(\boldsymbol{c}_t, \boldsymbol{q}_t)$ is the cosine similarity between the masked contextual representations produced before and after quantization. $\boldsymbol{\tilde{q}}$ is randomly sampled from $\boldsymbol{Q}_t$ which consists of $\boldsymbol{q}_t$ and ``Distractor'' labels from other masked time steps within the same speech utterance. $\kappa$ is the non-negative temperature parameter. 
The entropy-based diversity loss in the second term ensures the pre-training pseudo-labels cover all codebook entries equally.
$\alpha$ is a tuned weight.
$\bar {l}_{g,v}$ is the average logit of the $v$-th entry in the $g$-th codebook within a mini-batch.

\par
\noindent
\textbf{Fine-tuning}: Wav2vec2.0 is fine-tuned in a supervised mode using the Connectionist Temporal Classification (CTC) \cite{graves2006connectionist} loss. A randomly initialized linear layer is added on top of the context network to project the contextual representations into vocabulary tokens.

\vspace{-0.3cm}
\subsection{Pre-trained HuBERT Model}

The HuBERT \cite{hsu2021hubert} model pre-training alternates between two steps: 1) a clustering step to create pseudo-labels; and 2) a prediction step to produce labels for masked positions.
\par
\noindent
\textbf{Model Architecture}: The model architecture of HuBERT is similar to Wav2vec2.0, including a feature encoder, a $k$-means quantization module and a transformer-based context network followed by a projection layer.
\par
\noindent
\textbf{SSL Pseudo-labels}: The HuBERT model derives discrete speech units, denoted as $\boldsymbol{q}_t$, to serve as the pseudo-labels during pre-training. These pseudo-labels were produced using a total of $G$ distinct $k$-means clusters of varying codebook sizes. The clustering process is initially performed on MFCC features.
A latent speech representation $\boldsymbol{\hat s}_t$ is quantized into $G$ discrete units $\boldsymbol{q}^{1}_{t}, \boldsymbol{q}^{2}_t, \cdots, \boldsymbol{q}^{G}_{t}$. 
During the iterative refinement of pseudo-labels, the latent speech representation $\boldsymbol{\hat s}_t$ obtained from the context network at the $t$-th time step are then further $k$-means clustered to update the pre-training labels.
\par
\noindent
\textbf{SSL Criterion}: BERT style prediction of masked discrete speech units based on the following cross-entropy loss function is used in HuBERT pre-training,
\begin{equation}
\resizebox{0.9\linewidth}{!}{
\begin{math}
    \small
    \mathcal{L}_{mask}(\boldsymbol{Z}, \{\boldsymbol{Q}^{(g)} \}^{G}_{g=1}, {\cal{M}}) 
    = \sum_{t \in \cal{M}} \sum_{g=1}^{G} \log p\left(\boldsymbol{q}_t^{g} \mid \boldsymbol{{\bm Z^{(mask)}_t}} \right)
\vspace{-0.1cm}
\end{math}
}
\end{equation}
where $\cal M$ represents the position indices in masked pseudo-label prediction, $\boldsymbol{\bm Z^{(mask)}_t}$ denotes the partially masked version of the continuous speech representation $\boldsymbol{Z}$ at time step $t$.
$p(\boldsymbol{q}_t^{g} \mid \boldsymbol{\bm Z^{(mask)}_t)}$ is the probability of the discrete speech units of the $t$-th frame assigned by the $g$-th $k$-means model. 
\par
\noindent
\textbf{Fine-tuning}: During supervised fine-tuning using the CTC loss, the projection layer is replaced by a randomly initialized linear layer before the model parameters are updated.

\vspace{-0.4cm}
\subsection{Pre-trained WavLM Model}
WavLM \cite{chen2022wavlm} is also a masked prediction based SSL pre-trained model that shares the same model architecture, discrete speech representation based pre-training and fine-tuning procedures as HuBERT, except for the following two differences: \textbf{1)} the incorporation of gated relative position bias in the transformer self-attention module; and \textbf{2)} the use of cocktail party style mixture speech inputs simulated with multiple overlapping speakers and various background noises to produce more acoustic perturbation invariant speech representations.

\vspace{-0.4cm}
\subsection{Pre-trained Data2vec Model}
Data2vec \cite{pmlr-v162-baevski22a} is an SSL pre-trained model that shares the same architecture and fine-tuning procedure of Wav2vec2.0. It learns to predict latent speech representations of the complete input audio sequence given a partial view of such input.
\par
\noindent
\textbf{SSL Criterion}: The pre-training of Data2vec is performed using an exponential moving average in a teacher-student mode. The teacher parameterization is as \cite{mohamed2022self}:

\begin{equation}
    \small
    \setlength{\abovedisplayskip}{2pt}
    \setlength{\belowdisplayskip}{2pt}
    \boldsymbol{\theta}_{TM, i} = 
    \begin{cases}
        \boldsymbol{\theta}_{SM, 0} & \text{\textit{, i} = 0}\\
        \gamma \boldsymbol{\theta}_{SM, i} + (1 - \gamma)\boldsymbol{\theta}_{TM, i-1}& \text{\textit{, i} $>$ 0}
    \end{cases}
\end{equation}

\par
\noindent
where $\boldsymbol{\theta}_{TM, i}$ and $\boldsymbol{\theta}_{SM, i}$ denote the parameters of the teacher and student models at training step $i$ respectively. $\gamma$ is the exponential moving average decay rate.
For those masked time steps to be predicted by the student model, the training targets ${\bf y}_{t}$ are obtained using the average output of the top $K$ transformer blocks of the teacher model's context network. Let $\boldsymbol{\hat{c}}_t^{TM,l}$ denotes the normalized output of the $l$-th transformer block of the teacher model's context network at frame $t$, the student model's training targets are computed as ${\bf y}_{t} = \frac{1}{K}\sum_{l=L-K+1}^{L}\boldsymbol{\hat{c}}_t^{TM,l}$.
The regression loss between the targets and outputs of the Data2vec model is given by
\begin{equation}
    \small
    \setlength{\abovedisplayskip}{1pt}
    \setlength{\belowdisplayskip}{1pt}
    \mathcal L_{Data2vec, t} = 
    \begin{cases}
        \frac{1}{2} ({\bf y}_{t}-\boldsymbol{c}_{t}^{SM})^2 / \beta & \text{, $|{\bf y}_{t}-\boldsymbol{c}_{t}^{SM}| \leq \beta$}\\
        |{\bf y}_{t}-\boldsymbol{c}_{t}^{SM}| - \frac{1}{2} \beta& \text{, otherwise}
    \end{cases}
\end{equation}
\par
\noindent
where $\boldsymbol{c}_{t}^{SM}$ is the student model's predicted output at the $t$-th time step, and the threshold $\beta$ (e.g., 0.25 \cite{pmlr-v162-baevski22a}), controls the transition from a squared error based loss to an $L1$ loss.

\vspace{-0.1cm}
\section{SSL Features Based A2A Inversion}\label{sec:a2a_inversion}

Human speech production involves the coordinated movements of various articulators such as the lips, teeth, tongue and palate. Articulatory movement representations are inherently invariant to extrinsic acoustic distortion. They have been successfully applied to normal and pathological speech \cite{rudzicz2010articulatory, gonzalez2017direct, xiong2018deep, yilmaz2019articulatory} recognition. 
In practice, recording detailed articulatory movements and vocal tract shape normally requires the use of intrusive techniques such as electromagnetic articulography (EMA) 
or magnetic resonance imaging (MRI). 
Compared to EMA and MRI, ultrasound tongue imaging (UTI) \cite{stone2005guide, ribeiro2019speaker,cleland2019enabling} is more portable, non-invasive, and less costly. UTI utilizes B-mode diagnostic ultrasound to capture the tongue surface movements during speech production at a high frame rate \cite{ribeiro2021tal}. However, there are very few publicly available UTI corpora~\cite{cai2011recognition, eshky2019ultrasuite, ribeiro2021tal}, all of which are exclusively in English and limited in size. 
By far the largest Tongue and Lips (TaL) corpus \cite{ribeiro2021tal} contains 24 hours of parallel ultrasound, video, and audio data collected from 81 native English speakers.
\par
The scarcity of such specialist data hinders the practical and wider use of UTI-based articulatory features in ASR systems for normal, atypical speech task domains and across languages. An alternative approach to obtaining articulatory movement information is to estimate it from the more accessible acoustic speech signals. This requires neural network based acoustic-to-articulatory (A2A) inversion techniques \cite{papcun1992inferring, uria2012deep, xie2018investigation, maharana2021acoustic} to be used. As the A2A inversion model training only requires a subset of parallel acoustic-articulatory training materials, the resulting inversion model can be used to produce articulatory features when only speech audio recordings are available. A more extensive and practical application of articulatory features in ASR systems becomes feasible.

\begin{figure*}[htb]
\begin{minipage}[b]{1.0\linewidth}

  \centering
  \centerline{\includegraphics[width=17cm]{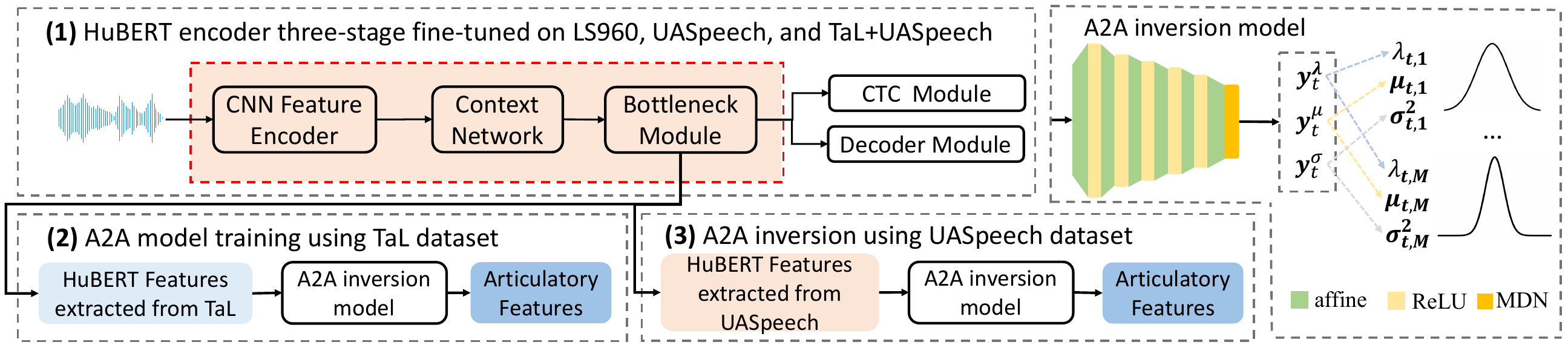}}

\end{minipage}
\vspace{-0.5cm}
\caption{An example of domain fine-tuned HuBERT feature based cross-domain acoustic-to-articulatory (A2A) inversion model architecture including: \textbf{(1)} the HuBERT encoder three-stage fine-tuned on out-of-domain 960-hour LibriSpeech, dysarthric UASpeech, and then the combined TaL+UASpeech audio data; \textbf{(2)} A2A model training using the HuBERT features extracted from TaL data and the parallel UTI-based articulatory features serving as the targets; \textbf{(3)} A2A inversion to generate UTI-based articulatory features using HuBERT features extracted from UASpeech.} 
\vspace{-0.4cm}
\end{figure*}

\par
To this end, in this paper the 24-hour out-of-domain non-aged healthy speech of the TaL dataset \cite{ribeiro2021tal} containing parallel UTI-based articulatory data is used to construct mixture density networks (MDN) based A2A inversion models \cite{hu2022exploiting} (the right lower part of Fig. 1). MDNs model the Gaussian mixture model density distribution parameters that characterize the articulatory movements instead of directly generating articulatory features. The MDN loss function is defined as
\begin{equation}
    \small
    \setlength{\abovedisplayskip}{1pt}
    \setlength{\belowdisplayskip}{1pt}
    \mathcal L_{MDN} = - \sum_{t=1}^{T} \ln \sum_{m=1}^M {\cal S}_m({\bf y}_{t}^{\bm {\lambda}}){\cal N}({\bf a}_{t};{\bm \mu}_{t,m}, {\bm \sigma}_{t,m}^{2})
\label{eq:mdn}
\end{equation}
\par
\noindent
where $M$ refers to the total number of mixture components, ${\bf a}_{t}$ is the UTI articulatory feature vector at the $t$-th frame, $\cal S$ and $\cal N$ respectively denote the Softmax activation and Gaussian distribution. As shown in Fig. 1, ${\bf y}_{t}^{\bm {\lambda}}$ represents the MDN network outputs fed into the Softmax activation to produce the mixture component weights ${\cal S}_m({\bf y}_{t}^{\bm {\lambda}})$ at time $t$. The mixture component mean and variance parameters at the $t$-th frame are predicted using the respective MDN outputs as ${\bm \mu}_{t,m} = {\bf y}_{t,m}^{\bm \mu}$, and ${\bm \sigma}_{t,m}^{2}= {\bf \exp}^{2}\left( {\bf y}_{t,m}^{\bm \sigma} \right)$. Following \cite{hu2022exploiting}, a multi-task learning (MTL) approach is adopted to construct the A2A inversion system. This A2A inversion system is trained using an interpolation between the MDN error cost of Eqn. \ref{eq:mdn}, the MSE loss and negative Pearson correlation criteria, all of which are computed against the ground truth UTI-based features:
\begin{equation}
    \small
    \setlength{\abovedisplayskip}{1pt}
    \setlength{\belowdisplayskip}{1pt}
    \mathcal L_{MSE} = \frac{1}{T} \sum_{t=1}^{T} ({\bf y}_{t} - {\bf a}_{t})^2
\label{eq:mse}
\end{equation}
\begin{equation}
    \small
    \setlength{\abovedisplayskip}{1pt}
    \setlength{\belowdisplayskip}{1pt}
    \mathcal L_{Pearson} = \frac{\sum_{t} ({\bf y}_{t} - \overline{{\bf y}})({\bf a}_{t} - \overline{{\bf a}})}{\sqrt{\sum_{t}({\bf y}_{t} - \overline{{\bf y}})^2}{\sqrt{\sum_{t}({\bf a}_{t} - \overline{{\bf a}})^2}}}
\label{eq:pearson}
\end{equation}
where ${\bf y}_{t}=\sum_{m}{\lambda_{t,m}\bm \mu}_{t,m}$ is the weighted sum of mixture component mean as the predicted articulatory features. $\overline{\bf y}$ and $\overline{\bf a}$ are respectively the average of predicted and ground true articulatory features over time $t$.
\par
Due to the large acoustic domain and language mismatch, a direct cross-domain and cross-lingual application of the A2A inversion model trained on non-aged and healthy acoustic-articulatory parallel data of the TaL corpus to the elderly and dysarthric speech is infeasible, as shown in the previous researches on cross-domain A2A inversion \cite{hu2022exploiting,hu2022exploitinguti}. To this end, such large acoustic domain and language mismatch can be minimized using either baseline multi-level adaptive networks (MLAN) \cite{liu2021recent, hu2022exploiting, hu2022exploitinguti} or domain fine-tuned SSL speech representations. Among these two approaches, the efficacy of MLAN domain adaptation is constrained by the data scarcity of both the source (TaL) and target (elderly and dysarthric) domains. In contrast, the SSL representations benefit from the additional use of large quantities of unlabelled data during their pre-training stage. Supervised fine-tuning is performed using combined datasets for each task. Take the UASpeech data for example, both the English TaL and UASpeech dysarthric speech were used in pre-trained HuBERT model fine-tuning. Similarly, for the Cantonese elderly speech data, both the English TaL and the Cantonese JCCOCC MoCA data were used during XLSR-128 model fine-tuning. The resulting SSL embeddings serve to produce more domain and language invariant representations compared with the original front-end features extracted from both datasets in the mix.
\par
The domain fine-tuned SSL speech features based cross-domain and cross-lingual adaptation considered in this paper is shown in Fig. 1, which includes the following steps: \textbf{a)} The encoder of SSL pre-trained speech model is three-stage fine-tuned on out-of-domain 960-hour LibriSpeech, dysarthric or elderly speech, and TaL speech plus dysarthric or elderly speech; \textbf{b)} the resulting fine-tuned model is then used to produce more domain and lingual invariant speech representations that exhibit smaller mismatch between these types of data. The resulting cross-domain and cross-lingual adapted speech representations are used in A2A inversion model training and articulatory feature generation for dysarthric or elderly speech data. The generated articulatory movement features are fused with the standard acoustic features via feature fusion to construct multi-modal TDNN or Conformer ASR systems (as the connections (c) and (d) shown in Fig. 2).

\vspace{-0.2cm}
\section{Pre-trained ASR Model Integration}\label{sec:sys_integrate}
\begin{figure*}[htbp]
    \centering
    \includegraphics[width=0.9\textwidth]{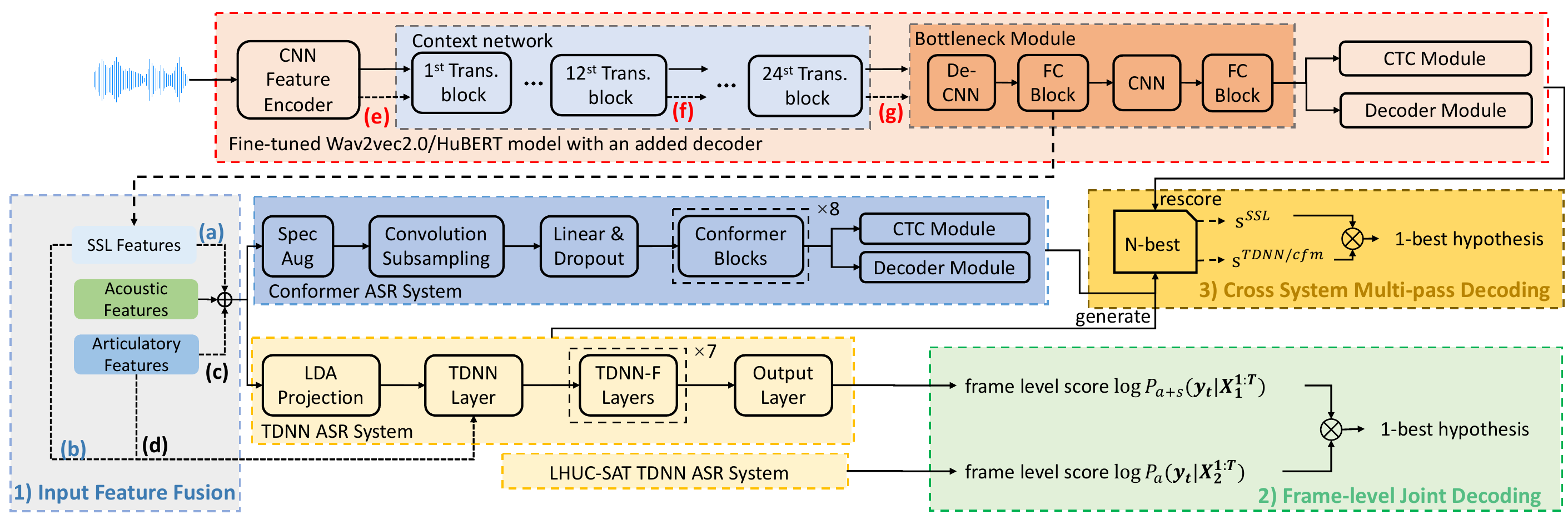}
    \vspace{-0.3cm}
    \caption[]{Dysarthric/elderly speech fine-tuned Wav2vec2.0/HuBERT models containing a ``Bottleneck Module'' (located at one of three different positions via \textcolor{red}{connections (e), (f) or (g)}) used to extract domain-adapted speech features. These models and their features are integrated into TDNN/Conformer ASR systems trained on in-domain dysarthric/elderly speech only using: \textbf{1)} \textcolor{SteelBlue}{input feature fusion} with standard acoustic frontends via \textcolor{SteelBlue}{connections (a) and (b)}; \textbf{2)} TDNN system \textcolor{MediumSeaGreen}{frame-level joint decoding} in the green box; and \textbf{3)} TDNN/Conformer systems' N-best outputs \textcolor{DarkGoldenrod}{multi-pass rescoring} using domain fine-tuned SSL pre-trained models in the brown box, as presented in Sec. \ref{sec:sys_integrate}. Connections (c) and (d) produce acoustic-articulatory speech recognition (AASR) systems using additional articulatory features predicted from domain-adapted SSL speech features via A2A inversion of Sec. \ref{sec:a2a_inversion}. ``Trans'' denotes transformer.
    }
    \label{fig:my_model}
    \vspace{-0.6cm}
\end{figure*}
Fundamental modeling differences exist between hybrid and end-to-end (E2E) ASR systems including current SSL pre-trained speech models. The conventional hybrid ASR architecture uses a modular design. It models acoustic, phonetic and language information separately. These are combined to produce the most likely word sequence during recognition. Model inference is performed in a frame-synchronous manner.
In contrast, a single neural network is used by E2E systems to directly convert the input sequence of frames into output labels, thus simplifying the overall system design. The latent alignment between the input frames and output labels is often learned using attention mechanisms, for example, in encoder-decoder based Conformer models. The resulting cross-system complementarity can be exploited using system combination. 
\par
In contrast to prior researches focusing on combining component hybrid and E2E systems that are trained using in-domain normal, dysarthric or elderly speech data only \cite{cui2022two, hu2022exploitinguti, liu2021recent, wang22k_interspeech}, this work aims to exploit the diversity and complementarity between SSL pre-trained speech models, and standard TDNN or Conformer ASR systems trained only on in-domain speech. 
To this end, a series of system integration approaches are explored in this paper.
\par
\noindent
\textbf{1) Input feature fusion} between standard acoustic front-ends, e.g., filter-banks, and domain-adapted SSL speech representations via either feature concatenation before being fed into the TDNN or Conformer system input layer, or optionally fusion at the TDNN hidden layer (connections (a) and (b) in Fig. 2).
\par
\noindent
\textbf{2) Time-synchronous frame-level joint decoding} is used to combine two or more hybrid TDNN systems \cite{swietojanski2013revisiting, liu2021recent, hu2022exploitinguti} which are separately trained using either standard acoustic features alone, or with additional fine-tuned SSL speech representations. A frame-level linear interpolation of system-specific acoustic log-likelihood scores (green dotted box in Fig. 2, bottom right) is used. Let $\log P_{a+s}(\bf{y}_t|\boldsymbol{X}_1^{1:\textit{\textbf{T}}})$ and $\log P_{a}(\bf{y}_t|\boldsymbol{X}_{2}^{1:\textit{\textbf{T}}})$ denote the acoustic log-likelihood scores at the $t$-th frame of TDNN systems constructed with or without additional SSL speech representations (marked as ``a+s'' and ``a''). The final frame-level score is obtained by
\begin{equation}
    \vspace{-0.1cm}
    \resizebox{0.9\linewidth}{!}{
    \begin{math}
     {\log \hat P(\bf{y}_t| \boldsymbol{X}_1^{1:\textit{\textbf{T}}}, \boldsymbol{X}_2^{1:\textit{\textbf{T}}})} = {\alpha \log P_{a+s}(\bf{y}_t|\boldsymbol{X}_1^{1:\textit{\textbf{T}}})} +
    {\beta \log P_{a} {(\bf{y}_{t} | \boldsymbol{X}^{1:\textit{\textbf{T}}}_{2})}}
\end{math}
}
\end{equation}
\par
\noindent
where $\alpha$ and $\beta$ are the empirically tuned system weights.
\par
\noindent
\textbf{3) Cross-system multi-pass decoding} involves a first decoding pass of TDNN or Conformer systems to produce the initial N-best outputs. These outputs are then rescored using domain-adapted SSL pre-trained models. Consider a sequence of acoustic features 
and its corresponding TDNN or Conformer (cfm) decoded N-best recognition hypotheses in \{$ Y_1, Y_2, ..., Y_N$\}. Let ${\textbf{s}}^{\text{TDNN/cfm}} = [s_1^{\text{TDNN/cfm}}, s_2^{\text{TDNN/cfm}}, ..., s_N^{\text{TDNN/cfm}}]^T$ and ${\textbf{s}}^{\text{SSL}} = [s_1^{\text{SSL}}, s_2^{\text{SSL}}, ..., s_N^{\text{SSL}}]^T$  respectively denote the N-best hypothesis score vectors produced by either the first-pass TDNN or Conformer systems and the second-pass SSL pre-trained models with fine-tuning. The final 1-best output is produced in the second-pass rescoring by applying fine-tuned SSL pre-trained models. During this rescoring stage, system-specific scores are interpolated for each N-best hypothesis as: 
\begin{equation}
    \setlength{\abovedisplayskip}{1pt}
    \setlength{\belowdisplayskip}{1pt}
    {\hat Y}_{best} = \mathop{\arg}\min_{i} \{\alpha s_i^{\text{SSL}} + \beta s_i^{\text{TDNN/cfm}}\}
\end{equation}
\par
\noindent
where $\alpha$ and $\beta$ are the weights assigned to the second-pass decoding by the fine-tuned SSL pre-trained model and the first-pass TDNN/Conformer systems,
as shown in Fig. 2 (right part of the middle, brown box).

\vspace{-0.3cm}
\section{ABLATION EXPERIMENTS ON IMPLEMENTATION DETAILS}\label{sec:implementation}

In this section, several key implementation issues affecting the performance of TDNN and Conformer ASR systems integrating a range of pre-trained models and their speech representations that are domain fine-tuned to dysarthric and elderly speech are discussed.
These include: \textbf{a)} the appropriate fine-tuning strategy and the choice of the most competitive baseline standalone SSL pre-trained models for dysarthric and elderly speech recognition; \textbf{b)} the detailed model structural configurations, and training costs used in these baseline pre-trained models; and \textbf{c)} the architectural modifications of these baseline pre-trained models that are required to extract suitable speech representations, and their fusion with standard acoustic features. A series of ablation studies are conducted on the UASpeech dysarthric speech corpus \cite{kim2008dysarthric} and the Dementiabank Pitt elderly speech corpus \cite{becker1994natural}. 

\begin{table*}[htbp]
\caption{Performance (WER\%) of pre-trained models that are fine-tuned either: \textbf{a)} on the 960-hour LibriSpeech or Common Voice out-of-domain normal speech data alone (``LS960'' or ``CV''); \textbf{b)} on the in-domain \textbf{UASpeech} dysarthric or \textbf{Dementiabank Pitt} elderly speech alone (``in-domain''); or \textbf{c)} two stages in turn respectively on the LibriSpeech or Common Voice data, and UASpeech or Dementiabank Pitt corpora. ``VL/L/M/H'' denote the ``Very Low'', ``Low'', ``Mild'' and ''High'' intelligibility subgroups in the UASpeech data. ``INV'' and ``PAR'' refer to the non-aged clinical investigator and the elderly participant in development (Dev.) and evaluation (Eval.) sets. ``All'' stands for the overall average WER.}
\centering
\vspace{-0.2cm}
\scalebox{0.9}{
\begin{tabular}{c|c|c|cc|cccc|c|cc|cc|c} \hline\hline
\multirow{3}{*}{Sys} & \multirow{3}{*}{model}              & \multirow{3}{*}{Fine-tune} & \multicolumn{7}{c|}{UASpeech WER (\%)}                                                                                                                      & \multicolumn{5}{c}{Dementiabank Pitt WER (\%)}                                              \\ \cline{4-15}
                     &                                     &                                     & \multirow{2}{*}{unseen} & \multirow{2}{*}{seen} & \multirow{2}{*}{VL} & \multirow{2}{*}{L} & \multirow{2}{*}{M} & \multirow{2}{*}{H} & \multirow{2}{*}{All} & \multicolumn{2}{c|}{Dev.} & \multicolumn{2}{c|}{Eval.} & \multirow{2}{*}{All}  \\ \cline{11-14}
                     &                                     &                                     &                         &                       &                     &                    &                    &                    &                      & PAR. & INV.               & PAR. & INV.                &                       \\ \hline\hline
$1$                  & \multirow{3}{*}{Wav2vec2.0}           & LS960 (A)                               &   -                      &  -                     &  94.51                   &  76.38                  &   51.51                 &  14.44                  &  54.59                    &  53.70    &  38.61                  &   44.43   &   42.40                  &    45.68                   \\
$2$                  &            & in-domain (B)                              &    79.99                     &  14.85                     &    63.81                 &   43.78                 &   35.65                 &  25.78                  &   40.40                   &  34.18    &    15.78                & 23.62     &    16.98                 &    24.42                   \\
$3$                  &                                     & LS960$\rightarrow$in-domain (C)                              &   55.75                      &  14.67                     &   62.76                  &   36.91                 &    25.24                &    9.11                &    30.78                  &  31.02    &  14.73                  & 21.18     &   14.76                  &   22.26                    \\ \hline
$4$                  & \multirow{3}{*}{\textbf{\begin{tabular}[c]{@{}c@{}}wav2vec2-\\conformer\end{tabular}}} & LS960 (A)                              &  -                       &   -             &  97.49                  &   83.95                 &   68.57                &  23.98               & 63.67                      &  56.10    &  39.22                  &  43.69    &  48.61                   &  46.95                    \\
$5$                  &  & in-domain (B)                              &   80.68                      &   15.48             &  63.69                   &  43.44                  &  37.69                  &  26.90                  &  41.05                    & 32.94     &  16.04                  & 22.02     & 15.43                    &  23.71                     \\
$6$                  &                                     & \textbf{LS960$\rightarrow$in-domain (C)}                               &   56.96                 &   14.14                    &  61.62                   &  36.33                  &  26.80                  &  9.85                  &    30.94                  &  29.71    &  14.29                  &  21.27    &  15.32                   &    \textbf{21.60}                   \\ \hline
$7$                  & \multirow{3}{*}{\textbf{HuBERT}}                              & LS960 (A)        &  -                &  -                           &  95.76                     &    81.31                 &  56.69                &    19.19              &   58.73                             & 53.01   &  36.55                &  41.42  &  40.07                 &  44.03                    \\
$8$                  &                               & in-domain (B)                              &   78.15                     &   13.69                 &   61.39                 &   41.61               &     35.10               &    25.05               &     38.97                &  32.87   &  15.43                 &  22.93   &  14.21                  &   23.55                    \\
$9$                  &                               & \textbf{LS960$\rightarrow$in-domain (C)}                               &   50.06                      &   13.30                    &   59.47                  &     33.62               &  22.22                  &   6.34                 &  \textbf{27.71}                    &   31.13   &   14.70                 &  21.94    &   14.65                  &  22.41                     \\ \hline
$10$                 & \multirow{3}{*}{\begin{tabular}[c]{@{}c@{}}Data2vec\\audio\end{tabular}}                     & LS960 (A)                              &   -                      &   -                   &  97.79                &   89.47                 &  72.76                &  36.35                &    70.14                 &   57.99    &   40.63                 &  44.95  &  46.84                  &  48.43                    \\ 
$11$                 &                      & in-domain (B)                               &   83.01                  &    14.47                &    62.80            &     43.22             &   37.86           &  28.41                 &       41.35           &  34.01     &   15.78              &  23.43   &  16.87                   &     24.31                  \\
$12$                 &                      & LS960$\rightarrow$in-domain (C)                               &   67.16                      &   14.42                    &   62.44                  &  41.33                  &  32.02                  &   14.87                 &   35.10                   &  33.50     &   15.96                 &  22.78    &  15.43                   &   24.03                    \\ \hline
$13$                  & \multirow{3}{*}{WavLM}                               & LS960 (A)        & -                    & -                   &  96.19              &  82.89                &  65.76            &  23.32                 &  62.36                &  50.33     &   35.40              & 40.35    &   39.96                  &  42.29                                      \\ 
$14$                 &                      & in-domain (B)    &  79.80             &  14.00                     &  61.35                   &   41.63                 &  34.73                  &   27.74                 &   39.80                   &  33.73    &   15.25                &  24.54     &  14.10                   &   24.09                                               \\
$15$                 &                      & LS960$\rightarrow$in-domain (C)                              &   52.47                  &   12.83                 &   59.02             &  33.87                &  23.57            &   7.62                &   28.38               &  30.12     &  14.76               &  20.76   &   14.43                  &   21.84                 \\ \hline
$16$                 &  \multirow{3}{*}{XLSR-53}                     & CV (A)                             &   -                  &  -                   &  95.60              &    85.82              &   64.22           &    30.91               &     65.26             &  60.11     &   46.76              &  49.80   &  49.50                   &   52.64                 \\
$17$                 &                     & in-domain (B)                             &   81.15                  &   14.42                 &    63.07            &   43.70               &    37.02          &     26.11              &   40.59            &  34.40   &  16.04      &  24.40      &  16.98                                &   24.74                 \\
$18$                 &                      & CV$\rightarrow$in-domain (C)                             &   63.26                  &   14.83                 &    62.78            &    39.74              &   29.24           &    13.67               &    33.82              &  33.28     & 16.10                & 23.83    &    15.87                 &   24.19                 \\
\hline\hline
\end{tabular}
}
\label{table_1}
\vspace{-0.4cm}
\end{table*}

\vspace{-0.2cm}
\subsection{Task Description}
\subsubsection{The English UASpeech Corpus}\label{subsubsec:ua} is the largest publicly available and widely used dysarthric speech dataset \cite{kim2008dysarthric}. It is an isolated word recognition task with $148912$ utterances and a vocabulary size of $455$, with approximately $103$ hours of speech from $29$ speakers, among whom $16$ are dysarthric speakers and $13$ are control healthy speakers. 
It is further split into 3 blocks, Block 1 (B1), Block 2 (B2), and Block 3 (B3) per speaker, each containing the same set of $155$ common words and a different set of $100$ uncommon words. 
After removing excessive silence at both the beginning and end of the speech audio segments using an HTK \cite{young2002htk} trained GMM-HMM system, a total of $30.6$ hours of audio data from B1 and B3 ($99195$ utterances) are used as the training set, while $9$ hours of dysarthric speech from B2 ($26520$ utterances) are used for performance evaluation.
Standard speaker-independent speed perturbation was used to expand the limited dysarthric speech (or elderly speech) training data using fixed perturbation factors \{0.9, 1, 1.1\}. Speaker-dependent speed perturbation was also used to modify the healthy, control speech data to that resembling the voice of each target dysarthric (or elderly) talker. For each dysarthric speaker, a perturbation factor is computed based on phonetic alignment analysis as described in \cite{xiong2019phonetic}. 
Such data augmentation produces a $130.1$ hours augmented training set ($399110$ utterances).

\subsubsection{The English TORGO Corpus \cite{rudzicz2012torgo}}\label{subsubsec:torgo} is a dysarthric speech dataset containing $8$ dysarthric and $7$ control healthy speakers with $13.5$ hours of speech ($16433$ utterances). Similar to the setting of UASpeech, a speaker-level data partition is conducted by combining all the $7$ control healthy speakers' data and two-thirds of the $8$ dysarthric speakers' data into the training set ($11.7$ hours). The remaining one-third of the dysarthric speech is used for performance evaluation ($1.8$ hours).
After the removal of excessive silence and data augmentation \cite{geng2020investigation, hu2022exploiting}, the augmented training and test sets respectively contain $34.1$ hours ($61813$ utterances) and $1$ hour ($1892$ utterances) of speech.
The entire TORGO dataset contains $1573$ distinct words.

\subsubsection{The English DementiaBank Pitt Corpus \cite{becker1994natural}}\label{subsubsec:dbank}
contains roughly $33$ hours of audio recorded from $292$ AD assessment interviews between elderly participants and clinical investigators. 
It is further divided into a $27.2$h training set, a $4.8$h development and a $1.1$h evaluation set.
The evaluation set contains the same $48$ speakers' Cookie Theft (picture description) task recordings as those in the ADReSS \cite{luz2020alzheimer} test set, while the development set covers the recordings of these speakers in other tasks, if available. 
The training set includes $688$ speakers ($244$ elderly participants and $444$ investigators), while the development and evaluation sets contain $119$ speakers ($43$ elderly participants and $76$ investigators) and $95$ speakers ($48$ elderly participants and $47$ investigators) respectively.
There is no overlapping between the elderly speakers in the training, development and evaluation sets.
After silence stripping \cite{ye2021development} and data augmentation via both speaker-independent and elderly speaker-dependent speed perturbation \cite{ye2021development}, the augmented training set contains $58.9$ hours of audio ($112830$ utterances) while the development and evaluation sets contain $2.5$ hours ($5103$ utterances) and $0.6$ hours ($928$ utterances) of audio, respectively.  

\subsubsection{The Cantonese JCCOCC MoCA Corpus} contains conversations from $256$ cognitive impairment assessment interviews between elderly participants and clinical investigators \cite{xuspeaker}. 
The training set includes $369$ speakers ($158$ elderly participants and $211$ investigators) with a duration of $32.4$ hours. 
The development and evaluation sets each contain speech from two different sets of $49$ elderly speakers not covered by the training set. 
After silence stripping and data augmentation similar to that used for the DementiaBank Pitt dataset, the augmented training set contains $156.9$ hours of audio ($389409$ utterances) while the development and evaluation sets contain $3.5$ hours ($13675$ utterances) and $3.4$ hours ($13414$ utterances) of audio, respectively. 

\begin{table}
\centering
\caption{Performance (CER\%) of pre-trained models fine-tuned directly on either the Common Voice out-of-domain normal speech data alone (``CV''), in-domain \textbf{JCCOCC MoCA} elderly speech alone (``in-domain''), or two stages respectively on the Common Voice, and JCCOCC MoCA datasets in turn.}
\centering
\vspace{-0.2cm}
\begin{tabular}{c|c|c|c|c|c} 
\hline\hline
Sys. & Model                 & Fine-tune     & DEV   & EVAL  & ALL    \\ 
\hline
1    & \multirow{3}{*}{XLSR-53} & CV           & 89.31 & 88.72 & 89.01  \\
2    &                       & in-domain    & 29.90 & 27.45 & 28.67  \\
3    &                       & CV$\rightarrow$in-domain & 31.47 & 29.21 & 30.33  \\ 
\hline
4    & \textbf{XLSR-128}                 & \textbf{in-domain}    & 28.86 & 26.53 & \textbf{27.69}  \\
\hline\hline
\end{tabular}
\label{table_1_jcmc}
\end{table}

\vspace{-0.2cm}
\subsection{Baseline Pre-trained ASR Model Fine-tuning}
\label{subsec:SSL models}

Two aspects of the baseline standalone pre-trained models are examined in the subsection.
\textbf{First}, the optimal strategy of supervised fine-tuning on limited dysarthric and elderly speech data based on either: \textbf{a)} a single-stage fine-tuning only on sufficient quantities of out-of-domain normal speech, for example, the 960-hour LibriSpeech data; \textbf{b)} a single-stage fine-tuning only on limited dysarthric or elderly speech; or \textbf{c)} a two-stage fine-tuning performed in turn on normal speech and in-domain dysarthric or elderly speech data.
\textbf{Second}, the precise SSL pre-trained model to choose among the following popular forms: Wav2vec2.0\footnote{https://huggingface.co/facebook/wav2vec2-large-960h-lv60}, Conformer based Wav2vec2.0 (wav2vec-conformer) with relative position
embeddings\footnote{https://huggingface.co/facebook/wav2vec2-conformer-rel-pos-large-960h-ft}, WavLM\footnote{https://huggingface.co/microsoft/wavlm-large}, HuBERT\footnote{https://huggingface.co/facebook/hubert-large-ls960-ft}, Data2vec audio model\footnote{https://huggingface.co/facebook/data2vec-audio-large-960h} and cross-lingual XLSR-53\footnote{https://huggingface.co/jonatasgrosman/wav2vec2-large-xlsr-53-english}.
\par
Several trends can be found in the results of Table \ref{table_1}. \textbf{1)} among all six pre-trained models, the above two-stage fine-tuning on the 960-hour LibriSpeech or Common Voice data first, followed by the in-domain UASpeech or DementiaBank Pitt data consistently produced the best performance across both types of data (Sys. 3, 6, 9, 12, 15, 18 \textit{vs.} the remaining systems in Table \ref{table_1}); 
\textbf{2)} for the UASpeech dysarthric speech data, the two-stage fine-tuned HuBERT model (Sys. 9, col. 10), which produced the lowest average WER of 27.71\%,
is adopted in the main experiments of the following Sec. \ref{sec:experiments} for dysarthric speech;
\textbf{3)} for the DementiaBank Pitt elderly speech data, the two-stage fine-tuned wav2vec-conformer model produced the lowest average WER of 21.60\% (Sys. 6, last col.), and is selected for the following experiments of Sec. \ref{sec:experiments} for the same task\footnote{Due to the much higher WER of multi-lingual XLSR-53 than English pre-trained models after two-stage fine-tuning (Sys. 18 \textit{vs.} Sys. 9 on UASpeech, Sys. 18 \textit{vs.} Sys. 6 on DementiaBank Pitt), no further experiments are conducted on the larger 128 languages based dataset trained multi-lingual XLSR-128 model for English dysarthric and elderly speech corpora.}.

\par

A comparable set of ablation studies are conducted on the Cantonese JCCOCC MoCA elderly speech dataset using the multi-lingual XLSR-53 model\footnote{https://huggingface.co/ctl/wav2vec2-large-xlsr-cantonese} as presented in Table \ref{table_1_jcmc} (Sys. 1-3)\footnote{Due to its language-specific characteristics, the Cantonese JCCOCC MoCA elderly speech dataset is different from all the other three English language based dysarthric and elderly speech corpora and requires separate selection of its baseline SSL pre-trained ASR model.}.
Due to a potential mismatch between the elderly JCCOCC MoCA corpus and Common Voice dataset, a single-stage direct fine-tuning on the in-domain  JCCOCC MoCA data produced the best overall performance (Sys. 2 \textit{vs.} Sys. 1, 3). A similar one-stage in-domain data fine-tuning experiment was further conducted on the 0.3B version of the multi-lingual XLSR-128 model \cite{babu2021xls} pre-trained on 128 languages. This model produced the lowest overall CER of 27.69\% among all fine-tuned systems (Sys. 4 \textit{vs.} Sys. 1-3) in Table \ref{table_1_jcmc} and is selected for the following experiments in Sec. \ref{sec:experiments}.

\vspace{-0.2cm}
\subsection{Improved 
 Model Architecture and Fine-tuning Criteria}
\label{subsec:decoder}
\begin{table*}[htbp]
\caption{Performance (WER\%) of pre-trained models (Sys. 9 of Table \ref{table_1}, HuBERT for UASpeech; Sys. 6 of Table \ref{table_1}, wav2vec2-conformer for Dementiabank Pitt) on the \textbf{UASpeech} and \textbf{Dementiabank Pitt} data with optionally added module and attention training loss.}
\vspace{-0.2cm}
\centering
\setlength\tabcolsep{3pt}
\begin{tabular}{c|c|c|c|cc|cccc|c|cc|cc|c} 
\hline\hline
\multirow{4}{*}{Sys} & \multirow{4}{*}{\begin{tabular}[c]{@{}c@{}}Encoder \\ Update\end{tabular}} & \multirow{4}{*}{\begin{tabular}[c]{@{}c@{}}Decoder \\(\#Transformer\\layers)\end{tabular}} & \multirow{4}{*}{\begin{tabular}[c]{@{}c@{}} Fine-tuning\\Loss\end{tabular}}     & \multicolumn{7}{c|}{UASpeech WER (\%)}                                                                                                                      & \multicolumn{5}{c}{Dementiabank Pitt WER (\%)}                                              \\ 
\cline{5-16}
                     &                          &                          &                                                                              & \multicolumn{7}{c|}{HuBERT (Sys. 9 in Tab. \ref{table_1})}                                                                                                              & \multicolumn{5}{c}{wav2vec2-conformer (Sys. 6 in Tab. \ref{table_1}))}                      \\ 
\cline{5-16}
                     &                          &                          &                                                                              & \multirow{2}{*}{unseen} & \multirow{2}{*}{seen} & \multirow{2}{*}{VL} & \multirow{2}{*}{L} & \multirow{2}{*}{M} & \multirow{2}{*}{H} & \multirow{2}{*}{All} & \multicolumn{2}{c|}{Dev.} & \multicolumn{2}{c|}{Eval.} & \multirow{2}{*}{All}  \\ 
\cline{12-15}
                     &                          &                          &                                                                              &                         &                       &                     &                    &                    &                    &                      & PAR.  & INV.              & PAR.  & INV.               &                       \\ 
\hline\hline
$1$                  & \multirow{4}{*}{\textbf{\begin{tabular}[c]{@{}c@{}}two-stage\\fine-tuning as \\ (C) in Table \ref{table_1}\end{tabular}}}                        & -                        & -                                                                            & 50.06                   & 13.30                 & 59.47               & 33.62              & 22.22              & 6.34               & 27.71                & 29.71 & 14.29             & 21.27 & 15.32              & 21.60                 \\ 
\cline{3-16}
$2$                &                          &  3                        &    \multirow{3}{*}{\textbf{\begin{tabular}[c]{@{}c@{}}attention\\(encoder frozen)\end{tabular}}}                                                                          & 47.79                   & 13.44                 & 58.40               & 32.55              & 20.39              & 6.50               & 26.91       & 29.02 & 13.77             & 20.28 & 14.21              & 20.91        \\
$3$                &                          &   \textbf{6}                       &                                                                              & 48.17                   & 13.49                 & 58.48               & 33.01              & 20.67              & 6.46               & \textbf{27.09}              & 28.85 & 13.44             & 19.84 & 13.87              & \textbf{20.63}        \\
$4$                &                          &   12                       &                                                                              & 49.19                   & 13.47                 & 58.97               & 33.30              & 21.41              & 6.67               & 27.48               & 28.61 & 13.37             & 19.97 & 14.21              & 20.53        \\
\hline\hline
$5$                & {\begin{tabular}[c]{@{}c@{}}+ a third stage\\joint fine-tuning\\ with decoder\end{tabular}} & 6       & {\begin{tabular}[c]{@{}c@{}}CTC:attention\\=3:7\end{tabular}}  & 50.50                   & 13.11                 & 58.95               & 33.16              & 22.12              & 7.26               & 27.78                & 29.26 & 13.36             & 18.75 & 12.99              & 20.55        \\
\hline\hline
\end{tabular}
\label{table_2}
\vspace{-0.4cm}
\end{table*}

In most previous researches, the ASR fine-tuning loss function of SSL pre-trained models is predominantly based on CTC \cite{graves2006connectionist}.
In order to improve the performance of pre-trained models during domain fine-tuning, the multi-task combination between CTC and attention based ASR costs \cite{kim2017joint, 9533587}, which are widely used in encoder-decoder based ASR systems \cite{gulati2020conformer}, are studied in this subsection. A number of further implementation issues appertaining to adjusting the best model complexity versus performance trade-off during pre-trained model fine-tuning are also investigated as follows: \textbf{a)} the number of transformer blocks in the decoder; \textbf{b)} the choice between either updating the decoder alone use the attention loss while freezing the pre-trained encoder, or jointly updating both the encoder and decoder using interpolated CTC and attention loss\footnote{the CTC and attention costs are linearly interpolated with a weighting of 3:7 following the ESPnet recipe in https://github.com/espnet/espnet/blob/master/egs/lrs2/asr1/conf and also our previous work \cite{geng2022speaker}.}.
The encoder parameters of the attention encoder-decoder (AED) model are initialized using the domain-adapted pre-trained models in Sec. \ref{subsec:SSL models} (i.e. HuBERT for UASpeech data shown as Sys. 9 and wav2vec2-conformer for Dementiabank Pitt corpus shown as Sys. 6 in Table \ref{table_1}, respectively).
The added decoder module is built using stacked transformer blocks with feedforward layer dimensionality set to 2048, each of which has 8 attention heads and an input dimensionality of 128.
\par
Several trends can be found in the results of Table \ref{table_2}: \textbf{1)} the fine-tuned pre-trained models with an added decoder module and its attention training cost outperformed those without such (Sys. 2-4 \textit{vs.} Sys. 1); 
\textbf{2)} using a decoder module containing 6 transformer layers, instead of 3 or 12, produced overall more balanced and competitive performance improvements over the comparable baseline fine-tuned HuBERT or wav2vec2-conformer models across both the UASpeech and Dementiabank Pitt tasks (Sys. 3 \textit{vs.} Sys. 2, 4), when separately fine-tuning the encoder and decoder in turn; 
\textbf{3)} further joint fine-tuning both the encoder and decoder on the in-domain data produced no consistent performance improvement (Sys. 5 \textit{vs.} Sys. 3). Based on these trends, the encoders of the HuBERT, wav2vec2-conformer models are two-stage fine-tuned\footnote{The XLSR-128 model's encoder is directly fine-tuned on the in-domain Cantonese elderly speech as in Sec. \ref{subsec:SSL models}.} using the CTC loss on the out-of-domain normal data first, followed by in-domain dysarthric or elderly speech data in turn as set out in Sec. \ref{subsec:SSL models}. The added decoder of 6 transformer layers is then fine-tuned using the attention loss on the in-domain data. Such settings are adopted in the main experiments of Sec. \ref{sec:experiments}.

\vspace{-0.2cm}
\subsection{SSL Speech Representation Extraction and Fusion}
\label{subsec:SSL extract}
The SSL pre-trained speech model can be used as a standalone speech recognition system after task fine-tuning. Alternatively, their speech representations can be incorporated into various back-end target domain data trained ASR systems via feature fusion based on, e.g., TDNN or Conformer, as considered in this paper. To this end, a ``Bottleneck Module'' (the upper part of Fig. 2, in the orange box) is introduced into the SSL pre-trained model to produce more compact speech representations. The ``Bottleneck module'' contains a stack of four interleaving convolutional and feed-forward layers: the first 1D transposed de-convolution CNN layer is used to change the stride length from 20 ms to 10 ms to allow a frame rate synchronization with that of the back-end ASR systems; a fully connected (FC) block, which consists of a linear layer, rectified linear unit (ReLU) activation and dropout module, is used to change the dimensionality of extracted features. This is followed by a CNN layer and a final FC block to revert the stride back to 20 ms and restore the dimensionality to 1024. The final SSL speech representations are extracted from the first FC layer.

\par
In this part of the ablation study, the effect of \textbf{1)} the form of feature fusion between SSL speech representations and front-end acoustic filter-bank (FBK) features; \textbf{2)} the dimensionality of extracted features (128, 256 and 512); and \textbf{3)} the position of the bottleneck module on the final system performance are analyzed. As shown in the left upper part of Fig. 2, three different positions of the bottleneck module (orange box in Fig. 2) are investigated: \textbf{(e)} after the CNN feature encoder and before the contextual transformer network; \textbf{(f)} immediately after the 12$^{th}$ transformer block inside the context network that contains a total of 24 transformer blocks; and \textbf{(g)} after the last, i.e. 24$^{th}$ transformer block of the context network.
\par
A consistent trend can be observed in the results presented in Table \ref{table_3}: \textbf{1)} combining the FBK features with fine-tuned HuBERT or wav2vec2-conformer features outperformed using the fine-tuned features alone on both dysarthric and elderly speech data (Sys. 2, 4, 6 \textit{vs.} Sys. 1, 3, 5); \textbf{2)} The performance comparison across different SSL speech feature dimensionality settings (128, 256 or 512) indicates that using 256 dimensions produced the overall best performance (Sys. 4 \textit{vs.} Sys. 2, 6);
\textbf{3)} Extracting fine-tuned HuBERT or wav2vec2-conformer features from the bottleneck module inserted immediately after the very last, 24$^{th}$ transformer block of the context network produced better performance than the other two locations (Sys. 4 \textit{vs.} Sys. 7, 8). Hence, the above settings are adopted in the main experiments of the following Sec. \ref{sec:experiments}.

\begin{table*}[htbp]
\caption{Performance (WER\%) of TDNN systems trained using fine-tuned HuBERT/wav2vec2-conformer features (feat.) extracted from bottleneck module located at varying positions inside Sys. 3 of Table \ref{table_2}, and different dimensionality (dim.) settings, with/without FBK features on 
\textbf{UASpeech} and \textbf{DementiaBank Pitt} data.}
\vspace{-0.2cm}
\centering
\setlength\tabcolsep{3pt}
\scalebox{0.9}{
\begin{tabular}{c|c|c|c|cc|cccc|c|cc|cc|c} 
\hline\hline
\multirow{4}{*}{Sys} & \multirow{4}{*}{\begin{tabular}[c]{@{}c@{}}bottleneck module\\position\end{tabular}}                               & \multirow{4}{*}{dim.}         & \multirow{4}{*}{features}        & \multicolumn{7}{c|}{UASpeech WER (\%)}                                                                                                                      & \multicolumn{5}{c}{DementiaBank Pitt WER (\%)}                                  \\ 
\cline{5-16}
                     &                                                                                                          &                               &                                  & \multicolumn{7}{c|}{HuBERT feat. (from Sys.3 in Tab. \ref{table_2})}                                                                                                   & \multicolumn{5}{c}{wav2vec2-conformer feat. (from Sys.3 in Tab. \ref{table_2})}           \\ 
\cline{5-16}
                     &                                                                                                          &                               &                                  & \multirow{2}{*}{unseen} & \multirow{2}{*}{seen} & \multirow{2}{*}{VL} & \multirow{2}{*}{L} & \multirow{2}{*}{M} & \multirow{2}{*}{H} & \multirow{2}{*}{All} & \multicolumn{2}{c|}{Dev.} & \multicolumn{2}{c|}{Eval.} & \multirow{2}{*}{All}  \\ 
\cline{12-15}
                     &                                                                                                          &                               &                                  &                         &                       &                     &                    &                    &                    &                      & PAR.  & INV.              & PAR.  & INV.               &                       \\ 
\hline\hline
$1$                  & \multirow{6}{*}{\begin{tabular}[c]{@{}c@{}}\textbf{after 24th}\\\textbf{transformer block}\end{tabular}} & \multirow{2}{*}{128}          & SSL feat.                & 53.79                   & 13.39                 & 57.45               & 32.85              & 23.39              & 12.03              & 29.23                & 27.95 & 13.61             & 19.40 & 13.76              & 20.26                 \\
$2$                  &                                                                                                          &                               & SSL feat. + FBK          & 52.96                   & 13.10                 & 56.95               & 31.91              & 21.75              & 12.53              & 28.73                & 27.46 & 13.28             & 19.53 & 12.76              & 19.93                 \\ 
\cline{1-1}\cline{3-16}
$3$                  &                                                                                                          & \multirow{2}{*}{\textbf{256}} & SSL feat.                & 53.14                   & 13.13                 & 56.92               & 32.06              & 22.55              & 12.26              & 28.82                & 27.73 & 13.22             & 19.34 & 13.43              & 20.00                 \\
$4$                  &                                                                                                          &                               & \textbf{SSL feat. + FBK} & 52.45                   & 12.99                 & 56.83               & 31.37              & 22.22              & 11.97              & \textbf{28.47}       & 27.21 & 13.20             & 19.13 & 12.87              & \textbf{19.73}        \\ 
\cline{1-1}\cline{3-16}
5                    &                                                                                                          & \multirow{2}{*}{512}          & SSL feat.                & 55.32                   & 13.10                 & 56.02               & 33.36              & 23.51              & 13.74              & 29.66                & 28.04 & 13.65             & 20.47 & 13.76              & 20.50                 \\
6                    &                                                                                                          &                               & SSL feat. + FBK          & 54.38                   & 13.18                 & 57.04               & 32.59              & 23.63              & 12.68              & 29.34                & 28.21 & 13.35             & 19.61 & 13.10              & 20.28                 \\ 
\hline
7                    & \begin{tabular}[c]{@{}c@{}}after 12th \\transformer block\end{tabular}                                   & 256                           & SSL feat. + FBK          & 52.07                   & 14.08                 & 56.02               & 31.39              & 22.90              & 13.58              & 28.97                & 35.52 & 15.39             & 27.06 & 14.65              & 25.29                 \\ 
\hline
8                    & \begin{tabular}[c]{@{}c@{}}after CNN \\feature encoder\end{tabular}                                      & 256                           & SSL feat. + FBK          & 53.40                   & 16.73                 & 61.48               & 33.33              & 23.96              & 14.40              & 31.11                & 45.09 & 18.16             & 33.68 & 16.87              & 31.38                 \\
\hline\hline
\end{tabular}
}
\label{table_3}
\vspace{-0.3cm}
\end{table*}

\vspace{-0.1cm}
\section{MAIN RESULTS}\label{sec:experiments}

The performance of the proposed methods to incorporate domain fine-tuned pre-trained speech models and their features into hybrid TDNN and Conformer ASR systems is examined in this section. 
Experiments are conducted on four datasets, the English UASpeech \cite{kim2008dysarthric} and TORGO \cite{rudzicz2012torgo} dysarthric speech corpora, as well as the English DementiaBank Pitt \cite{becker1994natural} and Cantonese JCCOCC MoCA \cite{xuspeaker} elderly speech datasets.
All pre-trained models use multi-stage fine-tuning and incorporate an additional decoder as described previously in Sec. \ref{subsec:SSL models} and \ref{subsec:decoder}. SSL speech representation extraction follows the bottleneck module design details of Sec. \ref{subsec:SSL extract}. 144-dimensional UTI-based articulatory features are extracted following our previous research \cite{hu2022exploitinguti}. Model-based speaker adaptation using learning hidden unit contributions (LHUC) \cite{swietojanski2016learning} is further applied to in-domain data trained TDNN systems.
\par

Sec. \ref{subsec:exp_dys} and \ref{subsec:exp_eld} provide details of the experiments conducted on two dysarthric speech corpora and two elderly speech datasets respectively.
Results are measured using WER for English corpora and CER for the Cantonese JCCOCC MoCA dataset. 
A matched pairs sentence-segment word error (MAPSSWE) based statistical significance test \cite{gillick1989some} at a significance level of $\alpha = 0.05$ is performed. 
In Sec. \ref{subsec:exp_ad}, speech recognition-based Alzheimer's disease (AD) detection is performed on the DementiaBank Pitt evaluation set (ADReSS2020 \cite{luz2020alzheimer}) using ASR  system outputs.

\vspace{-0.3cm}
\subsection{Experiments on Dysarthric Speech}\label{subsec:exp_dys}
\subsubsection{Baseline ASR System Description}
For the UASpeech dataset, hybrid LF-MMI factored time delay neural network (TDNN) systems \cite{peddinti2015time} containing $7$ context-slicing layers are trained following the Kaldi \cite{povey2011kaldi} chain system setup, except that i-Vector features are not incorporated. 
The E2E Conformer systems are implemented using the ESPNet toolkit\footnote{$8$ encoder + $4$ decoder layers, feed-forward layer dim = $1024$, attention heads = $4$, dim of attention heads = $256$, interpolated CTC+AED cost.} \cite{watanabe2018espnet} to directly model grapheme (letter) sequence outputs. 
$80$-dimensional FBK features are utilized in the Conformer systems, while $40$-dimensional FBK features and a $3$-frame context window are used in the hybrid TDNN system.
Following the configurations given in \cite{christensen2012comparative,christensen2013combining,sehgal2015model,yu2018development}, a uniform language model (LM) with a word grammar network is used in decoding. 

On the TORGO dataset, the hybrid LF-MMI TDNN and E2E graphemic Conformer systems use the same configurations as those adopted above for the UASpeech data, except that $40$-dimensional Mel-scale FBK features are used for both systems. 
A $3$-gram LM trained by all the TORGO transcripts with a vocabulary size of $1.6$k is used for both the TDNN and Conformer systems during recognition.

During the multi-stage fine-tuning of HuBERT as presented in Sec. \ref{subsec:SSL models}, the encoder is updated on the out-of-domain 960-hour LibriSpeech and then in-domain UASpeech/TORGO for 20 epochs data in turn. The added decoder is further fine-tuned for 10 epochs on the in-domain dysarthric speech data, while the encoder is frozen. A linearly decayed learning rate scheduling is used. The initial rate settings of 1e-5 and 1e-4 are used for the UASpeech and TORGO datasets, respectively.

\subsubsection{Performance Analysis on Dysarthric Speech}
Several trends are found among the UASpeech results of Table \ref{table_4}:

\begin{table*}
\centering
\caption{Performance of fine-tuned Hubert, TDNN or Conformer based ASR systems constructed with or without HuBERT features (``HuB Feat.''), and optionally using the cross-domain inverted UTI articulatory features on the dysarthric \textbf{UASpeech} and \textbf{TORGO} test sets;
``Seve./Mod.'' denote the ``Severe'' and ``Moderate'' intelligibility subgroups.
``MLAN'' is cross-domain multi-level adapted network \cite{hu2022exploitinguti} extracted bottleneck features. ``+'' and ``$\rightarrow$'' stand for frame-level joint decoding and multi-pass rescoring. $\dag$ and $\ast$ denote statistically significant differences (MAPSSWE \cite{gillick1989some}, $\alpha = 0.05$) obtained against the baseline HuBERT and Conformer systems (Sys. 4 and 12 for UASpeech, Sys. 5 and 12 for TORGO). ``$\downarrow^{X}$'' and ``$\Downarrow^{Y}$'' denote the absolute and relative WER reductions obtained over Sys. $X$ and $Y$ respectively.}
\label{table_4}
\vspace{-0.2cm}
\setlength\tabcolsep{2pt}
\centering
\scalebox{0.8}{
\begin{tabular}{c|c|c|c|cc|cccc|c|ccc|c} 
\hline\hline
\multirow{2}{*}{Sys} & \multirow{2}{*}{model}     & \multirow{2}{*}{\begin{tabular}[c]{@{}c@{}}acoustic \\ feature\end{tabular}} & \multirow{2}{*}{A2A input}             & \multicolumn{7}{c|}{UASpeech WER (\%)}                                                                                                                                & \multicolumn{4}{c}{TORGO WER (\%)}  \\ 
\cline{5-15}
                     &                            &                                                                              &                                        & unseen                & seen                  & VL                    & L                     & M                     & H                     & All                   & Seve. & Mod. & Mild & All           \\ 
\hline\hline
$1$                  & \multirow{2}{*}{TDNN}    & \multirow{2}{*}{FBK}                                                         & \xmark                  & 51.62                 & 16.98                 & 62.53                 & 31.92                 & 23.12                 & 13.67                 & 30.56                 & 12.80 & 8.78 & 3.64 & 9.47          \\
$2$                  &                            &                                                                              & MLAN \cite{hu2022exploitinguti}                                & 54.95                 & 17.64                 & 63.35                 & 33.14                 & 25.76                 & 15.78                 & 32.27                 & 12.80 & 4.59 & 2.71 & 8.35          \\ 
\hline
$3$                  & TDNN LHUC                      & FBK                                                                          & \xmark                  & 47.73                 & 15.49                 & 61.12                 & 28.95                 & 19.08                 & 11.94                 & 28.13                 & 12.52 & 8.27 & 3.25 & 9.11          \\ 
\hline\hline
$4$                  & HuBERT                     & \multirow{3}{*}{raw audio}                                                   & \multirow{3}{*}{\xmark} & 48.17($_{6.68}\Downarrow^{1}$)                 & 13.49($_{20.55}\Downarrow^{1}$)                 & 58.48($_{6.48}\Downarrow^{1}$)                 & 33.01                 & 20.67                 & 6.46($_{52.74}\Downarrow^{1}$)                  & 27.09($_{11.35}\Downarrow^{1}$)                 & 13.17 & 4.59 & 2.79 & 8.56          \\
\cline{2-2}\cline{5-15}
5                    & \begin{tabular}[c]{@{}c@{}}+in-domain\\3-gram LM\end{tabular}
                         &                                                                              &                                        & -                     & -                     & -                     & -                     & -                     & -                     & -                     & 12.03 & 4.59 & 2.79 & 7.97          \\ 
\hline
$6$                  & \multirow{2}{*}{TDNN}    & \multirow{2}{*}{FBK+HuB feat.}                                                     & \xmark                  & 52.45                 & 12.99                 & \textbf{56.83}$^\dag$ & \textbf{31.37}$^\dag$ & 22.22                 & 11.97                 & 28.47                 & 10.65 & 3.16 & 2.09 & 6.76$^\dag$          \\
$7$                  &                            &                                                                              & HuB feat.                              & 49.53                 & 13.08                 & \textbf{57.01}$^\dag$($_{1.47}\downarrow^{4}$) & \textbf{30.33}$^\dag$($_{2.68}\downarrow^{4}$) & 20.71                 & 10.29                 & 27.38                 & \textbf{9.72}  & \textbf{3.47} & \textbf{2.32} & \textbf{6.40}$^\dag$          \\ 
\hline
$8$                  & Sys.3 + 6                  & -                                                                            & -                                      &  40.37$^\dag$  & 11.71$^\dag$ & 52.46$^\dag$ & 24.59$^\dag$ & 15.33$^\dag$ & 7.48                  & 22.94$^\dag$($_{4.15}\downarrow^{4}$)  & 9.92  &  3.06 & 2.09 & 6.36          \\ 
$9$                  & Sys.3 + 7                  & -                                                                            & -                                     & 40.26$^\dag$   & 11.71$^\dag$ & 52.14$^\dag$ & 25.08$^\dag$ & 15.35$^\dag$ &  7.16             & 22.90$^\dag$($_{4.19}\downarrow^{4}$)  & 9.35  & 4.08 & 2.17 & 6.30          \\ 
$10$                  & Sys.3+6+7                  & -                                                                            & -                                      & \textbf{39.85}$^\dag$($_{8.32}\downarrow^{4}$) & \textbf{11.72}$^\dag$ & \textbf{52.07}$^\dag$($_{6.41}\downarrow^{4}$) & \textbf{24.69}$^\dag$ & \textbf{15.27}$^\dag$ & 7.10                  & \textbf{22.75}$^\dag$($_{4.34}\downarrow^{4}$) & 9.55  & 3.37 & 2.24 & \textbf{6.28}$^\dag$          \\ 
\hline
$11$                  & Sys.10$\rightarrow$4        & -                                                                            & -                                      & \textbf{34.28}$^\dag$ & \textbf{11.71}$^\dag$ & \textbf{50.70}$^\dag$ & \textbf{23.51}$^\dag$ & \textbf{12.06}$^\dag$ & \textbf{4.20}$^\dag$  & \textbf{20.56}$^\dag$($_{6.53}\downarrow^{4}$) & \textbf{9.11}  & \textbf{2.96} & 2.63 & \textbf{6.07}$^\dag$($_{1.90}\downarrow^{5}$)          \\ 
\hline\hline
$12$                 & \multirow{2}{*}{Conformer} & FBK                                                                          & \xmark                  & 99.30                 & 18.24                 & 66.77                 & 49.39                 & 46.47                 & 42.02                 & 50.03                 & 21.22 & 6.63 & 4.80 & 13.72         \\
$13$                 &                            & FBK+HuB feat.                                                                      & HuB feat.                              & \textbf{74.76$^\ast$} & \textbf{13.69$^\ast$} & \textbf{62.82$^\ast$} & \textbf{42.02$^\ast$} & \textbf{35.76$^\ast$} & \textbf{19.51$^\ast$} & \textbf{37.64$^\ast$}($_{12.39}\downarrow^{12}$) & \textbf{13.86} & \textbf{4.29} & \textbf{2.63} & \textbf{8.81$^\ast$}($_{4.91}\downarrow^{12}$)          \\ 
\hline
$14$                 & Sys.13~$\rightarrow$ 4     & -                                                                            & -                                      & \textbf{73.07$^\ast$} & \textbf{13.27$^\ast$} & \textbf{62.12$^\ast$} & \textbf{41.34$^\ast$} & \textbf{34.82$^\ast$} & \textbf{18.27$^\ast$} & \textbf{36.72$^\ast$} & \textbf{13.46} & \textbf{3.98} & \textbf{2.71} & \textbf{8.56$^\ast$}          \\
\hline\hline
\end{tabular}
}
\vspace{-0.5cm}
\end{table*}
\noindent
\textbf{i)} The standalone domain fine-tuned HuBERT model produced a statistically significant overall WER reduction of 3.47\% absolute (11.35\% relative) over the TDNN system using standard 40-dimensional FBK features (Sys. 4 \textit{vs.} Sys. 1).
\par
\noindent
\textbf{ii)} The \textbf{relative} WER reductions obtained using the fine-tuned HuBERT over the baseline TDNN system (Sys. 4 \textit{vs.} Sys. 1) were much larger for the seen words (20.55\%, col. 6) than the unseen words (6.68\%, col. 5). Such disparity in performance gains was also found between the very low (6.48\%, col.7, ``VL'') and high (52.74\%, col. 10, ``H'') intelligibility groups.
\par
\noindent
\textbf{iii)} Fusing the standard FBK front-ends with the fine-tuned HuBERT features (Fig. 2(a)-(b)), and further incorporating inverted UTI-based features (Fig. 2(c)-(d)) as TDNN inputs produced statistically significant WER reductions of 1.47\% and 2.68\% absolute (2.51\% and 8.12\% relative) over the standalone fine-tuned HuBERT model on the ``VL'' and ``L'' groups (Sys. 7 \textit{vs.} Sys. 4, col. 7, 8). 

\par
\noindent
\textbf{iv)} 2-way frame-level joint decoding between the LHUC speaker adapted TDNN system trained using the FBK features only, and that constructed using FBK plus HuBERT features (Sys. 3+6, shown as Sys. 8), or the one built also using the inverted UTI-based articulatory features (Sys. 3+7, shown as Sys. 9), consistently produced statistically significant overall WER reductions over the standalone HuBERT model (Sys. 8, 9 \textit{vs.} Sys. 4). The largest WER reduction of \textbf{4.34\%} absolute (\textbf{16.02\%} relative) was obtained using a 3-way frame-level joint decoding among all these three TDNN systems (Sys. 3+6+7, shown as Sys. 10) over the baseline HuBERT model (Sys. 10 \textit{vs.} Sys. 4)\footnote{System weights empirically set as 9:8, 7:9 for Sys. 8, 9 in 2-way joint decoding, and 8:5:5 for Sys. 10 in 3-way joint decoding.}. More specifically, statistically significant WER reductions of up to \textbf{8.32\%} and \textbf{6.41\%} absolute (\textbf{17.27\%} and \textbf{10.96\%} relative) on unseen words and ``VL'' group were respectively obtained (Sys. 10 \textit{vs.} Sys. 4, col. 5, 7).
\par
\noindent
\textbf{v)} Cross-system multi-pass rescoring the 3-way joint decoding's N-best (N=30) outputs using the standalone HuBERT model\footnote{The CTC, attention and TDNN system scores' weights were empirically set as 0.9:0.001:0.1 for Sys.11.} produced further performance improvements (Sys. 11 \textit{vs.} Sys. 10). A statistically significant overall WER reduction of \textbf{6.53\%} absolute (\textbf{24.10\%} relative) was obtained over the baseline fine-tuned HuBERT (Sys. 11 \textit{vs.} Sys. 4). An overall WER of \textbf{20.56\%} on the UASpeech test set (\textbf{Sys. 11}, \textbf{50.70\%} on "VL" group, \textbf{34.28\%} on unseen words) was obtained.

\par
\noindent
\textbf{vi)} Similar incorporating both the fine-tuned HuBERT features and the A2A inverted UTI features into the Conformer system produced a statistically significant overall WER reduction of \textbf{12.39\%} absolute (\textbf{24.77\%} relative) over the baseline Conformer using the FBK features alone (Sys. 13 \textit{vs.} Sys. 12). Further multi-pass rescoring using the fine-tuned HuBERT model produced marginal WER reductions (Sys. 14 \textit{vs.} Sys. 13), in contrast to the large improvements from HuBERT rescoring the TDNN systems' outputs (Sys. 11 \textit{vs.} Sys. 10). 
This is due to the larger model architecture similarity between HuBERT and Conformer models, in contrast to that between hybrid TDNN and HuBERT models. This limits their complementarity and improvements from system combination.
\par
Finally, the performance of our best system (\textbf{Sys.11, Table \ref{table_4}}) is compared with a set of state-of-the-art systems that are recently published on UASpeech task, and follow the same training-evaluation protocol\footnote{Block 1+3 data used in training, all the 16 dysarthric speakers of Block 2 for evaluation, and a 255 recognition vocabulary including both common and uncommon words \cite{christensen2012comparative, christensen2013combining, sehgal2015model, xiong2019phonetic, kim2008dysarthric}.}.
These are shown in Table \ref{table_5}. Our best system produced the lowest published overall WER of \textbf{20.56\%} (\textbf{50.70\%} on very low ``VL'' intelligibility).

\begin{table}[htbp]
\vspace{-0.3cm}
\centering
\caption{WER (\%) of recently published and ours on \textbf{UASpeech}.
}
\vspace{-0.2cm}
\scalebox{0.8}{
\begin{tabular}{c|c|c} 
\hline\hline
System                                                                   & VL      & All     \\ 
\hline\hline
Sheffield-2020 Fine-tuning CNN-TDNN speaker adaptation \cite{xiong2020source}                & 68.24 & 30.76  \\
CUHK-2021 NAS  DNN  +  Data  Aug.  +  LHUC-SAT  +  AV  fusion \cite{liu2021recent}        & 60.30 & 25.21  \\
CUHK-2022 DNN + Data Aug. + LHUC-SAT + AUV fusion \cite{hu2022exploiting}                    & 60.14 & 24.82  \\
CUHK-2022 DNN + Data Aug. + SBE Adapt + LHUC-SAT \cite{geng2022speaker}                             & 59.30 & 25.05  \\
CUHK-2022 TDNN + spectral basic GAN + LHUC-SAT \cite{jin2022personalized}                        & 59.18 & 27.85  \\
BUT-2022 Wav2vec2.0 + fMLLR + xvectors \cite{baskar2022speaker}                                & 57.72 & 22.83  \\
Nagoya Univ.-2022 WavLM \cite{violeta2022investigating} & 71.50 & 51.80 \\
FAU-2022 Cross-lingual XLRS + Conformer \cite{hernandez22_interspeech} & 62.00 & 26.10 \\
CUHK-2023 Kaldi TDNN + VAE-GAN + LHUC-SAT \cite{jin2023adversarial} & 57.31 & 27.78 \\
BTBU-2023 MAV-HuBERT + LRS3 visual fusion \cite{yu2023multi} & 63.98 & (27.94) \\ 
\textbf{TDNN + HuBERT fea. + sys. comb. (sys. 11 Tab. \ref{table_4}, ours) }& \textbf{50.70} & \textbf{20.56}  \\
\hline\hline
\end{tabular}
}
\label{table_5}
\vspace{-0.2cm}
\end{table}
A comparable set of experiments is conducted using the 34.1-hour speed perturbation augmented TORGO training dataset, as shown in Table \ref{table_4}. The following trends that are similar to those found on the UASpeech data are also observed:
\par
\noindent
\textbf{i)} 
The best-performing TDNN system utilizes the most powerful form of integration with the fine-tuned HuBERT model and features (Sys. 11). This system uses input feature fusion, HuBERT empowered A2A inverted UTI features, 3-way frame level joint decoding and multi-pass rescoring. It produced a statistically significant overall WER reduction of 1.90\% absolute (23.84\% relative) over the standalone fine-tuned HuBERT model supplemented with an external in-domain data constructed language model (Sys. 11 \textit{vs.} Sys. 5).
\par
\noindent
\textbf{ii)} Compared with the baseline Conformer system constructed using FBK features alone, incorporating both the HuBERT features, and the A2A inverted UTI articulatory features produced a statistically significant WER reduction of \textbf{4.91\%} absolute (\textbf{35.79\%} relative, Sys. 13 \textit{vs.} Sys. 12).

\vspace{-0.3cm}
\subsection{Experiments on Elderly Speech}\label{subsec:exp_eld}

\begin{table*}
\centering
\caption{Performance of fine-tuned pre-trained ASR models (``FT. pre-trained model'', wav2vec2-conformer for the English DementiaBank Pitt, XLSR-128 for the Cantonese JCCOCC MoCA), TDNN or Conformer based ASR systems constructed with or without domain-adapted SSL speech features (``SSL feat.''), and optionally using the cross-domain and cross-lingual inverted UTI features on the development (Dev) and evaluation (Eval) sets of \textbf{English DementiaBank Pitt} and \textbf{Cantonese JCCOCC MoCA} corpora. 
$\dag$ and $\ast$ denote statistically significant WER reductions (MAPSSWE \cite{gillick1989some}, $\alpha = 0.05$) obtained over the baseline wav2vec2-conformer/XLSR-128 and Conformer systems (Sys. 5 and 12). 
Other naming conventions follow Table \ref{table_4}.}
\label{table_7}
\vspace{-0.2cm}
\setlength\tabcolsep{3pt}
\begin{tabular}{c|c|c|c|cc|cc|c|cc|c} 
\hline\hline
\multirow{4}{*}{Sys} & \multirow{4}{*}{model}                                             & \multirow{4}{*}{\begin{tabular}[c]{@{}c@{}}acoustic \\ feature\end{tabular}} & \multirow{4}{*}{A2A input}             & \multicolumn{5}{c|}{DEMENTIABANK PITT WER (\%)}                                                                        & \multicolumn{3}{c}{JCCOCC~MOCA CER (\%)}                                                           \\ 
\cline{5-12}
                     &                                                                    &                                                                              &                                        & \multicolumn{5}{c|}{wav2vec2-conformer}                                                                               & \multicolumn{3}{c}{XLSR-128}                                                                      \\ 
\cline{5-12}
                     &                                                                    &                                                                              &                                        & \multicolumn{2}{c|}{Dev.}                     & \multicolumn{2}{c|}{Eval.}                    & \multirow{2}{*}{All}  & \multirow{2}{*}{Dev.}          & \multirow{2}{*}{Eval.}         & \multirow{2}{*}{All}            \\ 
\cline{5-8}
                     &                                                                    &                                                                              &                                        & ~PAR.                 & INV.                  & PAR.                  & INV.                  &                       &                                &                                &                                 \\ 
\hhline{============}
$1$                  & \multirow{2}{*}{TDNN}                                            & \multirow{2}{*}{FBK}                                                         & \xmark                  & 47.93                 & 19.91                 & 36.66                 & 19.76                 & 33.80                 & 26.87                          & 23.71                          & 25.28                           \\
$2$                  &                                                                    &                                                                              & MLAN \cite{hu2022exploitinguti}                                & 45.82                 & 19.21                 & 34.89                 & 18.42                 & 32.35                 & 25.06                          & 22.88                          & 23.96                           \\ 
\hline
$3$                  & TDNN LHUC                                                              & FBK                                                                          & \xmark                  & 45.49                 & 19.26                 & 35.44                 & 18.42                 & 32.33                 & 25.77                          & 22.94                          & 24.35                           \\ 
\hline\hline
$4$                  & \begin{tabular}[c]{@{}c@{}}FT. pre-trained model\end{tabular} & \multirow{2}{*}{raw audio}                                                   & \multirow{2}{*}{\xmark} & 28.85                 & 13.44                 & 19.84                 & 13.87                 & 20.63                 & 28.28                          & 26.17                          & 27.22                           \\
$5$                  & +in-domain 4-gram LM                                                                &                                                                              &                                        & 28.13                 & 13.06                 & 19.36                 & 13.87                 & 20.11($_{13.69}\downarrow^{1}$)                 & 26.41                          & 24.38                          & 25.39                           \\ 
\hline
$6$                  & \multirow{2}{*}{TDNN}                                            & \multirow{2}{*}{FBK+SSL feat.}                                                     & \xmark                  & 27.21          & 13.20                 & 19.13                 & 12.87                 & 19.73                 & 20.78\textbf{\textbf{$^\dag$}} & 18.21\textbf{\textbf{$^\dag$}} & 19.49\textbf{\textbf{$^\dag$}}  \\
$7$                  &                                                                    &                                                                              & SSL feat.                               & \textbf{26.43$^\dag$} & 12.77                 & \textbf{18.29$^\dag$} & 12.10                 & \textbf{19.08$^\dag$}($_{1.03}\downarrow^{5}$) & \textbf{20.01\textbf{$^\dag$}} & \textbf{17.69\textbf{$^\dag$}} & \textbf{18.84\textbf{$^\dag$}}  \\ 
\hline
$8$                 & Sys.3 + 6                                                          & -                                                                            & -                                      &  26.98$^\dag$ & 12.82 & 18.06$^\dag$ & 12.21                 & 19.29$^\dag$  &    19.08$^\dag$                       &    16.85$^\dag$                     &  17.96$^\dag$                           \\ 
$9$                 & Sys.3 + 7                                                          & -                                                                            & -                                      & 26.11$^\dag$ & 12.60 & 17.80$^\dag$ & \textbf{11.32}$^\dag$ & 18.78$^\dag$ &  18.94$^\dag$                         &   16.73$^\dag$                       &    17.83$^\dag$                       \\ 
$10$                 & Sys.3+6+7                                                          & -                                                                            & -                                      & \textbf{26.05$^\dag$} & \textbf{12.52}        & \textbf{17.59$^\dag$} & 11.43$^\dag$          & \textbf{18.69$^\dag$}($_{1.42}\downarrow^{5}$) & \textbf{18.82$^\dag$}                          & \textbf{16.58$^\dag$}                          & \textbf{17.69$^\dag$}                           \\ 
\hline
$11$                 & Sys.10$\rightarrow$4                                               & -                                                                            & -                                      & \textbf{25.27$^\dag$} & \textbf{12.07$^\dag$} & \textbf{16.73$^\dag$} & \textbf{11.88$^\dag$} & \textbf{18.07$^\dag$}($_{2.04}\downarrow^{5}$) & \textbf{18.57\textbf{$^\dag$}} & \textbf{16.28\textbf{$^\dag$}} & \textbf{17.42\textbf{$^\dag$}}($_{7.97}\downarrow^{5}$)  \\ 
\hline\hline
$12$                 & \multirow{2}{*}{Conformer}                                         & FBK                                                                          & \xmark                  & 48.71                 & 20.97                 & 36.93                 & 19.42                 & 34.57                 & 33.08                          & 31.24                          & 32.15                           \\
$13$                 &                                                                    & FBK+SSL feat.                                                                      & SSL feat.                               & \textbf{28.38$^\ast$} & \textbf{14.53$^\ast$} & \textbf{19.40$^\ast$} & \textbf{13.10$^\ast$} & \textbf{20.79$^\ast$}($_{13.78}\downarrow^{12}$) & \textbf{28.34\textbf{$^\ast$}} & \textbf{26.00\textbf{$^\ast$}} & \textbf{27.16\textbf{$^\ast$}}($_{4.99}\downarrow^{12}$)  \\ 
\hline
$14$                 & Sys.13$\rightarrow$4                                               & -                                                                            & -                                      & \textbf{27.66$^\ast$} & \textbf{13.96$^\ast$} & \textbf{18.81$^\ast$} & \textbf{12.87$^\ast$} & \textbf{20.16$^\ast$} & \textbf{28.19\textbf{$^\ast$}} & \textbf{25.88\textbf{$^\ast$}} & \textbf{27.03\textbf{$^\ast$}}  \\
\hline\hline
\end{tabular}
\vspace{-0.5cm}
\end{table*}

\subsubsection{Baseline ASR System Description} 
Following the Kaldi chain system setup, the hybrid TDNN system contains 14 context-slicing
layers with a 3-frame context. For all systems, 40-dimensional Mel-scale FBK features are utilized as input. On the English DementiaBank Pitt dataset, for both the hybrid TDNN and E2E graphemic Conformer systems\footnote{$12$ encoder + $12$ decoder layers, feed-forward layer dim = $2048$, attention heads = $4$, dim of attention heads = $256$, interpolated CTC+AED cost.}, a word-level $4$-gram language model (LM) with Kneser-Ney smoothing is trained using the SRILM toolkit \cite{stolcke02_icslp}. A $3.8$k word recognition vocabulary covering all words in the DementiaBank Pitt corpus is used during recognition. 
On the Cantonese JCCOCC MoCA data, the Conformer model training uses Cantonese characters as the output targets. 
A word-level $4$-gram language model with Kneser-Ney smoothing is trained on the transcription ($610$k words). A $5.2$k recognition vocabulary covering all the words in the JCCOCC MoCA corpus is employed.
\par
During the multi-stage fine-tuning of wav2vec2-conformer as presented in Sec. \ref{subsec:SSL models}, the encoder is initially fine-tuned on the out-of-domain 960-hr LibriSpeech and then in-domain DementiaBank Pitt corpus for 30 epochs in turn. For the JCCOCC MoCA dataset, the 0.3B version of XLSR-128 \cite{babu2021xls} is one-stage fine-tuned on in-domain elderly JCCOCC MoCA speech data only for 30 epochs. For both datasets, the added decoder is further fine-tuned for 10 epochs on the in-domain elderly speech data, while the encoder is frozen.

\subsubsection{Performance Analysis on Elderly Speech} Several trends are found among the DementiaBank Pitt results of Table \ref{table_7}:

\par
\noindent
\textbf{i)} Compared with the TDNN system using 40-dimensional FBK features, the standalone domain fine-tuned wav2vec2-conformer incorporating the in-domain data constructed 4-gram LM produced a large overall WER reduction of \textbf{13.69\%} absolute (\textbf{40.50\%} relative, Sys. 5 \textit{vs.} Sys. 1).
\par
\noindent
\textbf{ii)} The TDNN systems constructed by fusing the FBK and fine-tuned wav2vec2-conformer features, as well as the cross-domain inverted UTI-based articulatory features, produced a statistically significant overall WER reduction of 1.03\% absolute (5.12\% relative) over the standalone fine-tuned wav2vec2-conformer baseline model (Sys. 7 \textit{vs.} Sys. 5).
\par
\noindent
\textbf{iii)} 3-way frame-level joint decoding among the LHUC speaker adapted TDNN system trained on FBK features, and that constructed using FBK plus domain-adapted wav2vec2-conformer features, as well as that with additional inverted UTI articulatory features (Sys. 3+6+7, shown as Sys. 10)\footnote{3-way joint decoding's weights empirically set as 5:2:8 for Sys. 10.} produced a statistically significant overall WER reduction of \textbf{1.42\%} absolute (\textbf{7.06\%} relative) over the standalone domain-adapted wav2vec2-conformer model (Sys. 10 \textit{vs.} Sys. 5).
\par
\noindent
\textbf{iv)} Cross-system multi-pass rescoring the 3-way joint decoding's N-best (N=30) outputs using the standalone fine-tuned wav2vec2-conformer gave a statistically significant overall WER reduction of \textbf{2.04\%} absolute (\textbf{10.14\%} relative) over the baseline wav2vec2-conformer model (Sys. 11 \textit{vs.} Sys. 5)\footnote{The CTC, attention and TDNN system scores' weights were empirically set as 1:0.05:0.0075 for Sys. 11.}.

\par
\noindent
\textbf{v)} For Conformer systems, incorporating both the fine-tuned wav2vec2-conformer features and inverted UTI-based features produced a statistically significant overall WER reduction of \textbf{13.78\%} absolute (\textbf{39.86\%} relative) over the baseline FBK features trained Conformer model (Sys. 13 \textit{vs.} Sys. 12).
\par
A similar set of experiments are conducted on the Cantonese JCCOCC MoCA data, as shown in Table \ref{table_7}.  
Trends similar to those on the DementiaBank Pitt data are found:

\noindent
\textbf{i)} 
The best TDNN system using the most powerful form of integration with the fine-tuned XLSR-128 model and features (Sys. 11, through feature fusion, 3-way frame level joint decoding and multi-pass rescoring) gave a statistically significant overall CER reduction of \textbf{7.97\%} absolute (\textbf{31.39\%} relative) over the standalone fine-tuned XLSR-128 baseline with a 4-gram LM (Sys. 11 \textit{vs.} Sys. 5).
\par
\noindent
\textbf{ii)} Compared with the baseline Conformer using FBK features alone, incorporating both the XLSR-128 extracted, and inverted UTI features produced a statistically significant CER reduction of \textbf{4.99\%} absolute (\textbf{15.52\%} relative, Sys. 13 \textit{vs.} 12).

\vspace{-0.2cm}
\subsection{Experiments on AD Detection}\label{subsec:exp_ad}
In this subsection, the performance of speech recognition based Alzheimer's disease (AD) detection on DementiaBank Pitt evaluation set (ADReSS2020 \cite{luz2020alzheimer}) is evaluated using the speech transcripts obtained using various pre-trained ASR systems and features. Following our previous researches \cite{wang2023exploiting, wang22l_interspeech}, 768-dimensional outputs from the last hidden layer of BERT \cite{devlin2018bert}, or Roberta \cite{liu2019roberta} model, serve as a vector embedding to represent the speech content from each participant's AD assessment interview.
BERT and Roberta text encoders are fine-tuned on the 11591-word manual transcripts of the ADReSS training set with the Masked Language Modelling (MLM) and Next Sentence Prediction tasks using 15 different random seeds. Since the fine-tuning objective is based on the MLM cost instead of AD classification error, the detection accuracy scores measured at consecutive BERT or Roberta fine-tuning epochs in practice fluctuate. Models obtained at the final three update epochs during the 30-epoch fine-tuning were used to produce separate text embedding features for the back-end SVM-based AD classifiers. Hence, BERT and Roberta’s representations were used to construct separate AD classifiers and evaluated independently. Their respective AD detection outputs are combined by majority voting to reduce the risk of over-fitting and smooth the unstable performance. Mean, standard deviation and best AD detection scores are calculated.
\par
We then investigate the correlation between ASR performance and AD detection accuracy, and in particular, whether the elderly speech recognition performance improvements obtained in the experiments of Sec. \ref{subsec:exp_eld} will translate to higher AD detection accuracy. To this end, a diverse set of ASR systems are selected to generate the elderly adults’ speech transcripts for downstream text-based AD detection.
These include: 
\textbf{A)} the wav2vec2-conformer model without decoder one-stage fine-tuned on out-of-domain 960 hours of LibriSpeech (Sys. 4 in Table \ref{table_1}); 
\textbf{B)} the LHUC speaker adapted TDNN system trained using the FBK features of the in-domain DementiaBank Pitt data (Sys. 3 in Table \ref{table_7}); 
\textbf{C)} the standalone domain fine-tuned wav2vec2-conformer model with an external in-domain data constructed 4-gram language model (Sys. 5 in Table \ref{table_7}); 
\textbf{D)} N-best outputs produced by the 3-way TDNN joint decoding, which are further cross-system multi-pass rescored using the fine-tuned wav2vec2-conformer model
(Sys. 11 in Table \ref{table_7});
\textbf{E)} the standalone domain fine-tuned XLSR-53 model with an external in-domain data constructed 4-gram language model; and 
\textbf{F)} N-best outputs produced by the 3-way TDNN joint decoding, which are further cross-system multi-pass rescored using the fine-tuned XLSR-53 model (Sys. E and F are comparable to Sys. C and D except for replacing wav2vec2-conformer by XLSR-53).

\vspace{-0.2cm}
\begin{table}[htbp]
\centering
\caption{Accuracy (Acc.), sensitivity (Sen.), and specificity (Spec.) of AD detection measured in Mean, standard deviation (``Std'') and Best scores using ASR transcripts obtained using systems described in paragraph 2 of Sec. \ref{subsec:exp_ad}, as well as manual transcripts.}
\setlength\tabcolsep{1pt}
\centering
\scalebox{0.8}{
\begin{tabular}{c|c|c|ccc|ccc|ccc} 
\hline\hline
\multirow{2}{*}{Sys.} & \multirow{2}{*}{System}                                                                & WER (\%)  & \multicolumn{3}{c|}{Acc. (\%)}         & \multicolumn{3}{c|}{Sen. (\%)} & \multicolumn{3}{c}{Spec. (\%)}  \\ 
\cline{3-12}
                      &                                                                                          & Eval. PAR. & Mean           & Std  & Best           & Mean & Std & Best                                                                                    & Mean & Std & Best               \\ 
\hline\hline
1                     & \begin{tabular}[c]{@{}c@{}}Sys. \textbf{A} \\ (Sys.4 in Tab. \ref{table_1})\end{tabular}                          & 43.69     & 71.85          & 4.68 & 83.33          &  81.74    &  18.23   &  95.83                                                                                       & 60.31     & 17.74   &  100.00                  \\ 
\hline
2                     & \begin{tabular}[c]{@{}c@{}}Sys. \textbf{B} \\ (Sys.3 in Tab. \ref{table_7})\end{tabular}                          & 35.44     & 75.99          & 5.44 & 87.50          & 57.20     &  12.94   & 87.50                                                                                        &  94.78    &  4.99   &  100.00                  \\ 
\hline
3                     & \begin{tabular}[c]{@{}c@{}}Sys. \textbf{E} \\(XLSR-53 based, \\ akin to Sys. C)\end{tabular} & 21.42     & 80.59          & 5.28 & \textbf{97.92} &  73.41    & 8.39    &  95.83                                                                                       & 87.78     & 6.07    & 100.00                   \\ 
\hline
4                     & \begin{tabular}[c]{@{}c@{}}Sys. \textbf{C} \\ (Sys.5 in Tab. \ref{table_7})\end{tabular}   & 19.36     & 80.61          & 4.77 & 91.67          &  76.28    & 9.14    &  95.83                                                                                       &  84.94    &  6.11   &   95.83                 \\ 
\hline
5                     & \begin{tabular}[c]{@{}c@{}}Sys. \textbf{F} \\(XLSR-53 based, \\ akin to Sys. D)\end{tabular} & 19.08     & \textbf{83.94} & 4.58 & \textbf{95.83} &  \textbf{84.56}    &  6.26   &  \textbf{100.00}                                                                                       &  83.31    & 7.42    &   100.00                 \\ 
\hline
6                     & \begin{tabular}[c]{@{}c@{}}Sys. \textbf{D} \\ (Sys.11 in Tab. \ref{table_7})\end{tabular}                        & 16.73     & \textbf{83.23} & 4.73 & \textbf{95.83} & \textbf{81.92}     & 6.78    &  \textbf{100.00}                                                                                       &  82.40    &  6.69  &  95.83                 \\
\hline\hline
7                     & Manual                         & -    &  81.09 & 3.95 & 91.67 & 70.37     & 5.73    &   83.33                                                                                      &  91.81   & 4.78   &  100.00                \\
\hline\hline
\end{tabular}
}
\label{table_9}
\vspace{-0.2cm}
\end{table}

Several trends can be found in the AD detection performance presented in Table \ref{table_9}:
\textbf{1)} In general, there is a correlation between the WER of ASR systems and the accuracy of AD systems: the lower the WER, the higher the AD detection performance (Sys. 3-6 \textit{vs.} Sys. 1,2). 
\textbf{2)} While the TDNN systems integrating the fine-tuned wav2vec2-conformer or XLSR-53 and their features are originally designed to optimize the ASR performance, they also produce more balanced and consistent AD detection accuracy performance (measured in terms of the mean accuracy scores) than the standalone pre-trained models after domain fine-tuning (Sys. 5, 6 \textit{vs.} Sys. 3, 4).
\textbf{3)} In addition, such improvements in AD detection accuracy are also consistent with those over the sensitivity scores of each system shown in the same table. For ASR-based automated AD detection systems, higher sensitivity scores can reduce the percentage of false negative diagnoses.
\textbf{4)} The state-of-the-art mean AD detection accuracy of \textbf{83.94\%} (standard deviation of 4.58\%, best score of 95.83\%) was obtained on the ADReSS20 test set consisting of 48 elderly speakers (Sys. 5). The best score of AD detection accuracy from our system is contrasted against recently published results on ADReSS2020 in Table \ref{table_10}.

\begin{table}
\vspace{-0.2cm}
\centering
\caption{AD detection accuracy (\%) on the \textbf{ADReSS2020} \cite{luz2020alzheimer} data between published systems and our system.}
\label{table_10}
\vspace{-0.2cm}
\scalebox{0.8}{
\begin{tabular}{c|c|c} 
\hline\hline
Sys.                        & Modality                       & Best AD Acc. (\%)  \\ 
\hline\hline
Ye et al. \cite{ye2021development}                           & Text (TDNN ASR)                      & 87.5  \\
Wang et al. \cite{wang22k_interspeech}                           & Text (Conformer ASR)                      & 91.7  \\
Syed et al. \cite{syed2021automated}                           & Text(manual)                   & 91.7  \\
Li et al. \cite{li2021comparative}                           & Text (TDNN ASR)                     & 88.0  \\
Martinc et al. \cite{martinc2021temporal}                           & Audio + Text (Manual)          & 93.8  \\
Laguarta et al. \cite{laguarta2021longitudinal}                           & Audio + pre-trained biomarkers & 93.8  \\
Wang et al. \cite{wang2023exploiting}                           & Text (CNN-TDNN ASR)                     & 93.8  \\
\textbf{ours (Sys. 5 in Table \ref{table_9})} & \textbf{Text (TDNN + SSL ASR)}                     & \textbf{95.8}  \\
\hline\hline
\end{tabular}
}
\vspace{-0.4cm}
\end{table}

\vspace{-0.3cm}
\section{DISCUSSION AND CONCLUSIONS}\label{sec:discusion_conclusion}
Our work has revealed the performance fragility of current SSL pre-trained speech foundation models when being applied to dysarthric and elderly speech data that are highly scarce, mismatched and diverse. Their performance disparity is found in several practical scenarios: a) between seen and unseen words in the often very limited dysarthric speech; b) between impaired speakers with high and very low speech intelligibility; and c) between non-aged clinical investigators and elderly adults, for example, in the DementiaBank Pitt data. To mitigate the above performance disparity issues, this paper explores a series of approaches to integrating cross-domain adapted SSL pre-trained foundation models and their features into TDNN and Conformer ASR systems for dysarthric and elderly speech recognition. Domain-adapted SSL speech representations are further utilized in acoustic-to-articulatory (A2A) inversion to construct multi-modal dysarthric and elderly ASR systems. 
Experiments conducted on four dysarthric or elderly speech datasets across two languages suggest that the proposed SSL pre-trained model and feature integration approaches can effectively improve the final ASR system's generalization performance on the highly scarce and mismatched dysarthric and elderly speech data. 
The TDNN systems constructed by integrating domain-adapted HuBERT, wav2vec2-conformer or multi-lingual XLSR models and their features consistently outperform the standalone fine-tuned SSL models by statistically significant WER or CER reductions of \textbf{6.53\%}, \textbf{1.90\%}, \textbf{2.04\%} and \textbf{7.97\%} absolute (\textbf{24.10\%}, \textbf{23.84\%}, \textbf{10.14\%} and \textbf{31.39\%} relative) on the four tasks respectively. The lowest published WERs of \textbf{20.56\%} (\textbf{50.70\%} on very low intelligibility, \textbf{34.28\%} on unseen words) and \textbf{18.07\%} are obtained on the benchmark UASpeech test set of 16 dysarthric speakers and the DementiaBank Pitt evaluation set of 48 elderly subjects respectively. 
Future research will focus on the rapid personalization of pre-trained ASR models for diverse dysarthric and elderly speakers, and effective pre-trained model compression approaches to improve their efficiency for practical deployment.
\vspace{-0.1cm}
\section*{Acknowledgment}

This research is supported by Hong Kong RGC GRF grant No. 14200220, 14200021, TRS T45-407/19N, Innovation \& Technology Fund grant No. ITS/218/21, 
National Natural Science Foundation of China 62106255, Research Project of Institute of Software, Chinese Academy of Sciences (ISCAS-ZD-202401, ISCAS-JCMS-202306) and Youth Innovation Promotion Association CAS Grant 2023119.

\bibliographystyle{IEEEtran}
\bibliography{ref}

\end{document}